\documentclass[3p,times,onecolumn,numbers]{elsarticle}
\usepackage{xcolor, lineno}
\usepackage{amssymb}
\usepackage{amsmath}
\usepackage{booktabs}
\usepackage{array}
\usepackage{graphicx}
\usepackage[colorlinks=true]{hyperref}
\usepackage{booktabs}   
\usepackage{threeparttable}
\usepackage{graphicx}  
\usepackage{xcolor}
\usepackage{colortbl}
\usepackage{multirow}
\modulolinenumbers[1]
\usepackage{booktabs}
\usepackage{array}
\journal{Materials Chemistry and Physics}

\begin{document}

\begin{frontmatter}

\title{Linking Electronic Bonding and Short-range Order to Strength in $\alpha$-Titanium Alloys: A First-Principles Study}
\author{Md Faiz Akhtar}
\author{Nilesh P. Gurao}
\author{Somnath Bhowmick\corref{cor1}}
\ead{bsomnath@iitk.ac.in}
\cortext[cor1]{Corresponding author}

\affiliation{Department of Materials Science and Engineering, Indian Institute of Technology Kanpur, Kanpur 208016, UP, India}

\begin{abstract}
The development of accurate strength prediction models for titanium alloys is critical for advanced materials design. This study systematically examines how the mechanical properties of $\alpha$-Ti are affected by substitutional (X = Al, V, Mo) and interstitial (Y = H, C, N, O) alloying elements, with a focus on electronic bonding. Using density functional theory (DFT), we uncover the short-range ordering (SRO) of substitutional atoms and quantify their influence on the electronic bonding and mechanical behavior. The primary novelty of this work lies in developing a predictive model for tensile strength that goes beyond traditional empirical approaches. To quantify the contributions of individual solutes to strengthening, we use physically grounded quantum-chemical descriptors, such as the Integrated Crystal Orbital Hamilton Population (ICOHP), which is a direct measure of bond strength derived from first-principles calculations. The resulting formula quantitatively predicts the tensile strength of a wide range of $\alpha$-Ti alloys, demonstrating a significant advancement in the computational design of high-performance structural materials.
\end{abstract}



\end{frontmatter}

\section{Introduction}
Titanium alloys occupy a distinctive niche among structural materials, offering an exceptional combination of high specific strength, corrosion resistance, biocompatibility, and moderate high-temperature stability. These attributes underpin their widespread use in aerospace, biomedical, and chemical processing applications. In contrast, pure $\alpha$-titanium, characterized by its hexagonal close-packed (HCP) crystal structure, exhibits relatively modest mechanical performance. Commercial grades typically achieve tensile strengths of 210–550 MPa. Moreover, the limited number of independent slip systems, basal and prismatic $\langle a\rangle$ slip with critical resolved shear stresses of approximately 209 and 181 MPa, respectively, and pyramidal $\langle c{+}a\rangle$ slip near 474 MPa, leads to pronounced anisotropy, restricted ductility, and poor formability. Consequently, alloying is not merely advantageous but fundamentally necessary to overcome these intrinsic constraints.

Substitutional element (X) additions are the principal levers of solid-solution strengthening in $\alpha$-Ti, acting through lattice distortion, modulus contrast, and dislocation-core modification~\citep{feng2023microstructure,kolli2018review, williams2020opportunities}. For example, Ti-6Al-4V accounts for roughly 50\% of global titanium consumption~\citep {boyer1996overview}. Interstitial element (Y) additions, such as H, C, N, and O, further modulate properties. Oxygen, for instance, raises yield strength to ${\sim}1$~GPa at 0.3~wt.\% at cryogenic temperature, but at the cost of ductility and fracture toughness~\citep{chong2020mechanistic, yu2015origin, faiz2025novel}. In Ti-X-Y systems, the interplay between substitutional and interstitial species introduces configurational complexity that classical treatments struggle to capture.

A critical yet frequently overlooked aspect of this complexity is chemical short-range ordering (SRO). High-resolution experimental techniques, such as energy-filtered TEM, atom-probe tomography, and diffuse scattering, together with DFT pair-interaction calculations, consistently reveal that substitutional solutes in $\alpha$-Ti deviate strongly from a random solid-solution distribution at the scale of a few nearest-neighbor shells~\citep{zhang2019direct, li2023quantitative, he2024quantifying}. For Al in $\alpha$-Ti, both DFT and experimental studies document a pronounced preference for Al-Al separation, driven by elastic strain relief and Fermi-level hybridization changes~\citep{zhang2019direct, kwasniak2023competition, yin2017comprehensive}. This ordering measurably elevates the critical resolved shear stress and yet is entirely absent from most predictive strength models. Systematic DFT studies of SRO in Ti-V and Ti-Mo remain comparatively rare~\citep{wu2025observation, bakulin2022impurity}, and the substantially different size and electronic structures of V and Mo relative to Al mean that Ti-Al SRO behavior cannot simply be transferred to these systems~\citep{lindwall2018diffusion, calazans2024recent}.

The introduction of interstitial solutes into binary Ti-X systems creates Ti-X-Y alloys with a richer SRO landscape. Interstitials occupy octahedral sites in the HCP lattice, but whether those sites are Ti-only octahedral voids or mixed Ti-X octahedral voids influences local chemical order and mechanical response~\citep{faiz2025novel, ghazisaeidi2014interaction, hooshmand2018first}. DFT work on Ti-Al-O by Gunda \textit{et al.} demonstrated that oxygen preferentially avoids Al-rich coordination shells owing to unfavorable O-Al electronic interactions~\citep{chong2020mechanistic, bakulin2022impurity, gunda2020understanding}. However, this particular study did not account for the SRO identified by Kwasniak \textit{et al.}~\citep{kwasniak2023competition}. Very few analogous investigations for V, Mo, and non-oxygen interstitials (H, C, N) are available in the literature~\citep{faiz2025novel, ghosh2021effect, ren2025effect}.

Against this backdrop, quantitative strength-prediction strategies for $\alpha$-Ti alloys broadly fall into two categories. Semi-empirical approaches based on the Empirical Electron Theory (EET) relate valence-electron descriptors, such as covalent electron number, bond length, and lattice electron density, to experimentally measured strengths~\citep{huang2023strengthening, qu2011theoretical, lin2011analysis}. Although computationally efficient, EET is constrained by several fundamental limitations. Its bond-energy terms are empirically parameterized rather than derived from first principles, which restricts transferability across alloy chemistries. The framework further assumes random solid solutions, thereby neglecting bond statistics arising from short-range order (SRO). This leads to physically inconsistent interpretations, for example, prior EET analyses of Ti–Al systems attribute strengthening primarily to Al–Al bonds, despite their low statistical likelihood at dilute Al concentrations and low thermodynamic likelihood  due to SRO, as atomic configurations with Al-Al bonds are less favorable~\citep{kwasniak2023competition, huang2023strengthening}. These deficiencies become more pronounced in ternary systems containing interstitials, where strongly directional $p$–$d$ covalent interactions between C, N, or O and Ti cannot be adequately captured within a metallic-bond formalism~\citep{qu2011theoretical, lin2016simple}.

First-principles density functional theory (DFT) offers a parameter-free alternative, and the Crystal Orbital Hamilton Population (COHP), as implemented in the LOBSTER package, provides a rigorous framework for quantifying bond strength via the Integrated COHP (ICOHP)~\citep{dronskowski1993crystal, maintz2016lobster, deringer2011crystal}. By decomposing the DFT band-structure energy into bond-resolved contributions, ICOHP directly links electronic structure to bonding, with more negative values indicating stronger interactions, and does so without recourse to empirical parameters~\citep{rohling2019correlations, steinberg2018crystal}. Although COHP/ICOHP has been widely applied to qualitative bonding analysis and thermodynamic stability studies~\citep{faiz2025novel, das2024interplay, al2024accelerating}, it has not yet been leveraged to develop explicit, composition-dependent tensile-strength relations for titanium alloys. Addressing this gap forms the central motivation of the present work.

Here we present a comprehensive first-principles investigation that addresses this gap. We systematically characterize short-range order (SRO) in binary Ti–X systems (X = Al, V, Mo), and further examine the role of interstitial solutes (Y = H, C, N, O) using DFT. This analysis is complemented by detailed electronic-structure characterization, including projected density of states (pDOS), $-p$COHP, ICOHP, integrated crystal orbital bond index (ICOBI), and molecular-orbital representations, to establish quantitative bonding descriptors and a mechanistic framework. Together, these efforts culminate in the development and experimental validation of an ICOHP-based tensile-strength model that explicitly incorporates SRO, replacing empirically assigned valence parameters with quantum-mechanically derived bond-strength descriptors.


\section{Computational Methodology}
\subsection{DFT Details}
Density Functional Theory (DFT) calculations were performed using a plane-wave basis set (with a kinetic energy cutoff of 500 eV) and Perdew-Burke-Ernzerhof (PBE)-Generalized Gradient Approximation (GGA) for the exchange-correlation functional, as implemented in the Vienna Ab-initio Simulation Package (VASP), using the Projector-Augmented-Wave (PAW) method for describing the interaction between the ionic cores and the valence electrons~\citep{PhysRevB.59.1758, PhysRevB.54.11169, PhysRevLett.77.3865,blochl1994projector}. The valence electron configurations used are: Ti (4s, 4p, 3d), Al (3s, 3p), V (4s, 4p, 3d), Mo (5s, 5p, 4d), H (1s), C (2s, 2p), N (2s, 2p), and O (2s, 2p). The Brillouin zone is sampled using a Monkhorst-Pack k-point mesh~\citep{monkhorst1976special}, ensuring convergence of the total energy to within 1 meV/atom. All atomic structures are fully relaxed until the Hellmann-Feynman forces on each atom are less than 0.01 eV/\AA, and the total energy change between successive self-consistent field iterations is less than 10$^{-8}$ eV. More details on the simulation parameters, such as KPOINTS, supercell size, etc., are provided in Table S1, Section S1 of the Supplementary Material (SM). Calculated lattice parameters and elastic constants C$_{ij}$ of pure $\alpha$-Ti (Table S2 of the SM) agree well with reported values, which validates the computational parameters used in this work.

\textcolor{black}{All calculations correspond to $T = 0$~K and zero external pressure. Pure $\alpha$-Ti and every alloy supercell (binary Ti-X, binary Ti-Y, and ternary Ti-X-Y) were fully relaxed with respect to cell volume, cell shape, and internal atomic coordinates, until all Hellmann-Feynman forces fell below 0.01~eV/\AA. The equilibrium lattice parameters of pure $\alpha$-Ti are listed in Table~S2 of the SM.}

\subsection{SRO: Binary Ti-X Systems}
\begin{figure}
\centering
\includegraphics[width=0.65\linewidth]{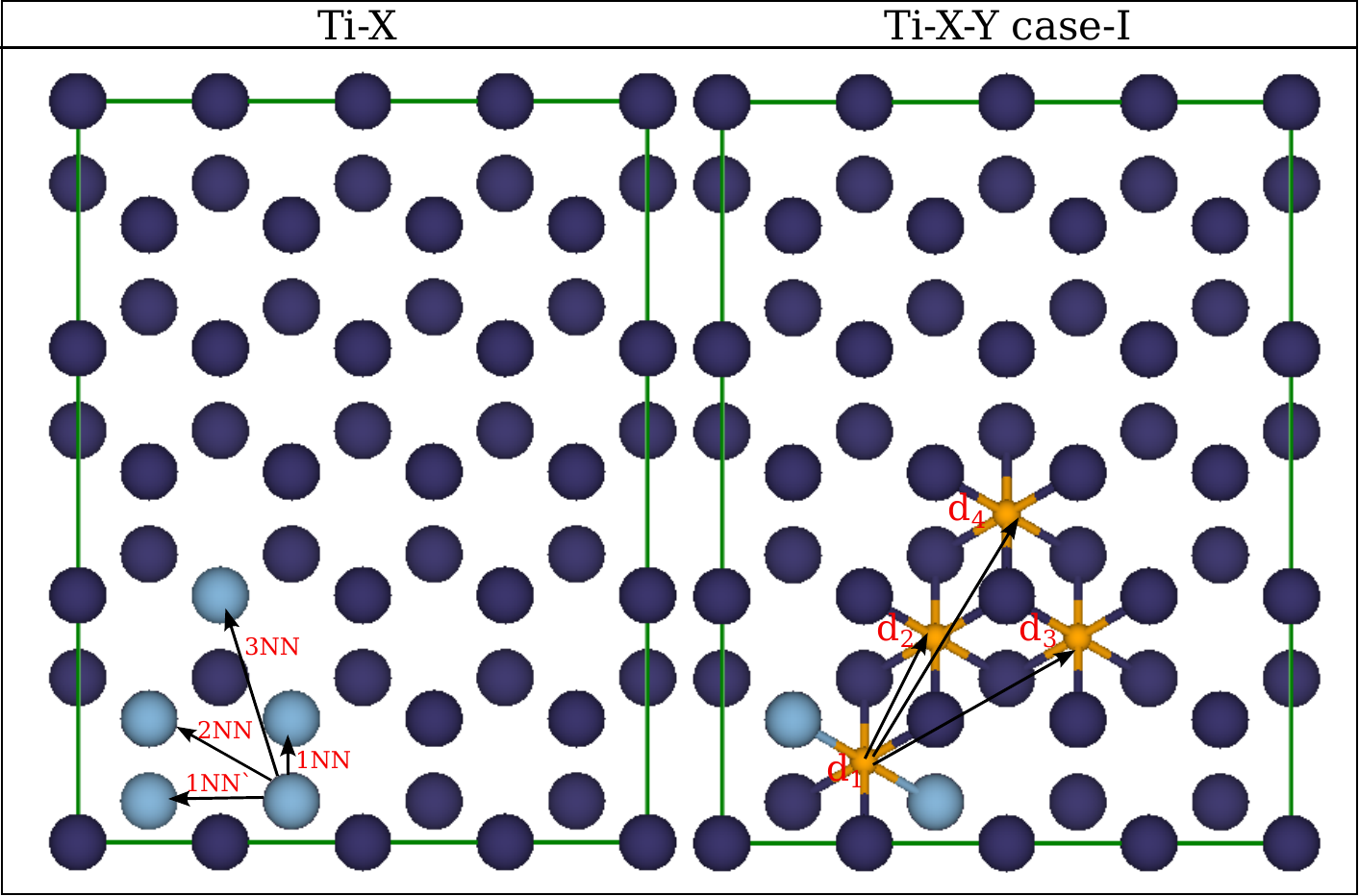}
\caption{The 144-atom supercell used for investigating Short-Range Ordering (SRO) in $\alpha$-Ti. (Column 1) Binary Ti-X, in which one substitutional atom (sky-blue) is fixed and the second is placed at varying distances (1NN to 3NN). (Column 2) A fixed substitutional SRO and an interstitial atom (orange) located at variable distances.}
\label{fig:schematic_sro}
\end{figure}
Short-range ordering (SRO) analysis in binary Ti-X alloys needed a large supercell containing 144 atoms, ensuring minimum image distances exceeding $\approx$12, 15, and 14 ~\AA~ in x, y, and z direction, which is sufficient to eliminate artificial interactions arising from periodic boundary conditions while maintaining computational tractability. For binary Ti-X systems, the binding energy was calculated using~\citep{kwasniak2023competition}:
\begin{equation}
    E_b = 2E_{\text{Ti+X}} - E_{\text{Ti}} - E_{\text{Ti+2X}}
    \label{eq:Eb_binary}
\end{equation}
where $E_{\text{Ti}}$ represents the total energy of the pure 144-atom $\alpha$-Ti, $E_{\text{Ti+X}}$ is the total energy of the supercell containing one substitutional X atom (143 Ti atoms + 1 X atom), and $E_{\text{Ti+2X}}$ is the total energy of the supercell containing two substitutional X atoms (142 Ti atoms + 2 X atoms). $E_b$ quantifies the energetic preference for two solute atoms (X) to occupy specific relative positions in the titanium lattice, with positive values indicating attractive interactions (thermodynamic driving force for SRO) and negative values indicating repulsive interactions.

The calculation procedure systematically varied the separation between the two X atoms from first-nearest-neighbor (1NN) to sixth-nearest-neighbor (6NN) positions in the HCP structure. One X atom was fixed at a reference lattice site, while the second X atom was placed sequentially at distinct nearest-neighbor positions, as shown in column 1 of Fig.~\ref{fig:schematic_sro}. For each configuration, the structure was fully relaxed, and the binding energy was calculated from the total energy of the supercell using Eq.~\ref{eq:Eb_binary}. The specific nearest-neighbor distances investigated were: 1NN out of plane and in plane at approximately 2.86 and 2.92~\AA, respectively; 2NN at approximately 4.08~\AA~(between adjacent basal planes); 3NN at approximately 5.09~\AA; 4NN at approximately 5.47~\AA, 5NN at approximately 6.5~\AA, and 6NN at approximately 7.7~\AA. 


\subsection{SRO: Ternary Ti-X-Y Systems}

The investigation of ternary Ti-X-Y systems, incorporating both substitutional (X~=~Al, V, Mo) and interstitial (Y~=~H, C, N, O) alloying elements, requires consideration of additional configurational degrees of freedom beyond those in binary systems. Specifically, interstitial atoms can occupy octahedral sites  surrounded by a mixture of Ti and X atoms, or they can occupy an octahedral site surrounded exclusively by Ti atoms. To map this energy landscape, the two substitutional X atoms were placed at their energetically favorable 2NN SRO positions, as determined from binary Ti-X calculations, and one interstitial Y atom was introduced at an octahedral site with mixed Ti-X coordination (Fig.~\ref{fig:schematic_sro}). The interstitial atom was then systematically displaced to successive nearest-neighbor octahedral sites ($d_1$ through $d_4$), and the binding energy was calculated for each configuration. This approach directly probes whether the pre-existing substitutional SRO cluster attracts or repels the interstitial, and identifies the thermodynamically preferred interstitial site in the ternary system.

The binding energy $E_b$ for ternary systems was calculated as:
\begin{equation}
E_b = E_{\text{Ti+2X}} + E_{\text{Ti+Y}} - E_{\text{Ti}} 
      - E_{\text{Ti+2X+Y}}
\label{eq:Eb_ternary}
\end{equation}
where $E_{\text{Ti}}$ is the total energy of pure $\alpha$-Ti, $E_{\text{Ti+2X}}$ is the total energy with two substitutional X atoms at their 2NN SRO configuration, $E_{\text{Ti+Y}}$ is the total energy with one interstitial Y atom at an octahedral site surrounded entirely by Ti atoms, and $E_{\text{Ti+2X+Y}}$ is the total energy of the ternary system with both substitutional and interstitial atoms present in the specific configuration being evaluated. This formulation isolates the interaction energy between the substitutional and interstitial subsystems, revealing whether their combined presence results 
in synergistic (positive $E_b$) or antagonistic (negative $E_b$) effects relative to their independent dissolution in the $\alpha$-Ti lattice.

\subsection{Electronic Structure: DOS}
The electronic density of states (DOS) provides fundamental information about the distribution of electronic energy levels, helping to understand materials' properties. Total DOS and atom-projected DOS (pDOS) were calculated using the tetrahedron method with Bl\"ochl corrections for high accuracy~\citep{blochl1994improved}. The pDOS decomposes the total DOS into contributions from specific atoms and their orbital angular momentum channels (s, p, d character), revealing which atomic orbitals are contributing at different energy ranges. 


\subsection{Electronic Structure: COHP}
Crystal orbital Hamilton population (COHP) analysis was performed using the LOBSTER package~\citep{maintz2016lobster} to quantify the nature of chemical bonding beyond DOS analysis. LOBSTER projects the plane-wave DFT wavefunctions from VASP onto local atomic orbital basis sets, enabling the partitioning of the band-structure energy into contributions from specific atom pairs and the reconstruction of a chemically intuitive picture of bonding in extended solids. The COHP for an atomic pair $i-j$ is defined as:
\begin{equation}
\text{COHP}_{ij}(E) = -\sum_{k,n} H_{ij}^{kn}(E) \rho_{ij}^{kn}(E) \delta(E - E_{kn})
\end{equation}
where $H_{ij}^{kn}$ is the Hamilton matrix element between orbitals of $i^{th}$ and $j^{th}$ atoms for $n^{th}$ band at $k^{th}$ reciprocal lattice point, $\rho_{ij}^{kn}$ is the density matrix element, and $E_{kn}$ are the Kohn-Sham eigenvalues. The $-$pCOHP (negative of pCOHP) is conventionally plotted as a function of energy, with positive values indicating bonding character and negative values indicating anti-bonding character.

The integrated $-p$COHP (ICOHP) provides a single scalar measure of bond strength, given by,
\begin{equation}
\text{ICOHP}_{ij} = \int_{-\infty}^{E_F} -\text{pCOHP}_{ij}(E) \, dE.
\end{equation}
Note that ICOHP has units of energy (eV) and represents the contribution of the specific $i$-$j$ bond to the total band-structure energy, making it a physically meaningful bonding descriptor directly derived from DFT calculations. More negative ICOHP values indicate stronger bonds. In the present study, ICOHP values were calculated for all relevant bond types: Ti-Ti, Ti-X, Ti-Y, X-Y, and X-X bonds. Average ICOHP values were computed for bond types occurring multiple times in the supercell. \textcolor{black}{The local orbital basis sets used for the LOBSTER projection are listed in Table~S1 of the SM.}


\subsection{Electronic Structure: ICOBI \textcolor{black}{and Population Analysis}}
The integrated crystal orbital bond index (ICOBI) provides a measure analogous to the bond order in molecular quantum chemistry, quantifying the covalent character of bonds~\citep{muller2021crystal}. ICOBI values range from 0 (no covalent bonding) to approximately 1 for single bonds, with higher values for multiple bonds. The ICOBI for pair $i$-$j$ is calculated as:
\begin{equation}
\text{ICOBI}_{ij} = \int_{-\infty}^{E_F} \text{COBI}_{ij}(E) \, dE,
\end{equation}
where COBI is the crystal orbital bond index.

\textcolor{black}{ICOBI quantifies the covalent component of a bond, with larger values indicating a higher effective bond order. A bonding scenario cannot, however, be assigned from ICOBI alone, since it carries no information about polarity. Following Reitz \textit{et al.}~\citep{reitz2024bonding}, we therefore combine it with a population analysis. Mulliken and L\"owdin charges were obtained from the same LOBSTER projection and are used jointly with ICOBI. Being basis-set dependent, the charges are used only as relative trends. The molecular orbitals in Figs.~\ref{fig:TiX_dos_cohp_mo} and~\ref{fig:Ti_Y_cohp_dos_MO} were obtained from the same projection using the linear combination of fragment orbitals (LCFO) scheme of LOBSTER 5.1.0~\citep{muller2024fragment}, taking each Ti-X or Ti-Y pair as a fragment, and were rendered in VESTA~\citep{momma2011vesta}.}

\subsection{GSFE}
Generalized stacking fault energy or the $\gamma$-surface is an essential energetic descriptor for understanding slip resistance in hcp metals~\cite{vitek1968intrinsic,lindaprb}. In this work, GSFE calculations were performed for the substitutional Ti-X systems (X = Al, V, Mo) using the same theoretical framework as in our earlier study on Ti with interstitial solutes~\cite{faiz2025novel}. We examine the basal $\{0002\}\langle 11\bar{2}0\rangle$ and prismatic $\{10\bar{1}0\}\langle 11\bar{2}0\rangle$ slip systems, as they represent the primary deformation modes in $\alpha$-Ti. For each slip system, supercells were constructed with lattice vectors aligned to the respective fault planes. All computational parameters, including the $ k$-point mesh and supercell size, are provided in Table~S1 of SM. 

The GSFE curves were generated by rigidly shifting the upper half of the crystal along the Burgers vector $\vec{b}=a/3[11\bar{2}0]$ in incremental steps from $0$ to $1\,\vec{b}$. At each displacement, atomic relaxation was allowed only along the direction normal to the fault plane, while in-plane coordinates and cell vectors are fixed to avoid artificial normal stresses. The stacking fault energy was computed as,
\begin{equation}
\gamma(\vec{u})=\frac{E(\vec{u})-E_0}{A},
\label{eq:gsfe}
\end{equation}
where $E(\vec{u})$ denotes the relaxed total energy at a given displacement $\vec{u}$, $E_0$ is the reference energy of the undeformed cell, and $A$ is the projected area of the fault plane.

In Ti-X alloys, chemical short-range ordering (SRO) favors X-X separation at the second-nearest-neighbor position. Accordingly, three configurations were constructed: FP, where the two X atoms lie directly above and below the fault plane; FP1, where the fault plane is one atomic layer below the X-X pair; and FP2, where it is two layers below. These configurations capture the variation in the local chemical environment at different distances from the fault plane. Similar configurations are considered for Ti-Y systems (Y = H, C, N, O) ~\cite{faiz2025novel}.

\section{Results and Discussion}
\subsection{SRO in Binary Ti-X and Ternary Ti-X-Y Systems}
\begin{figure}
\centering
\includegraphics[width=0.8\linewidth]{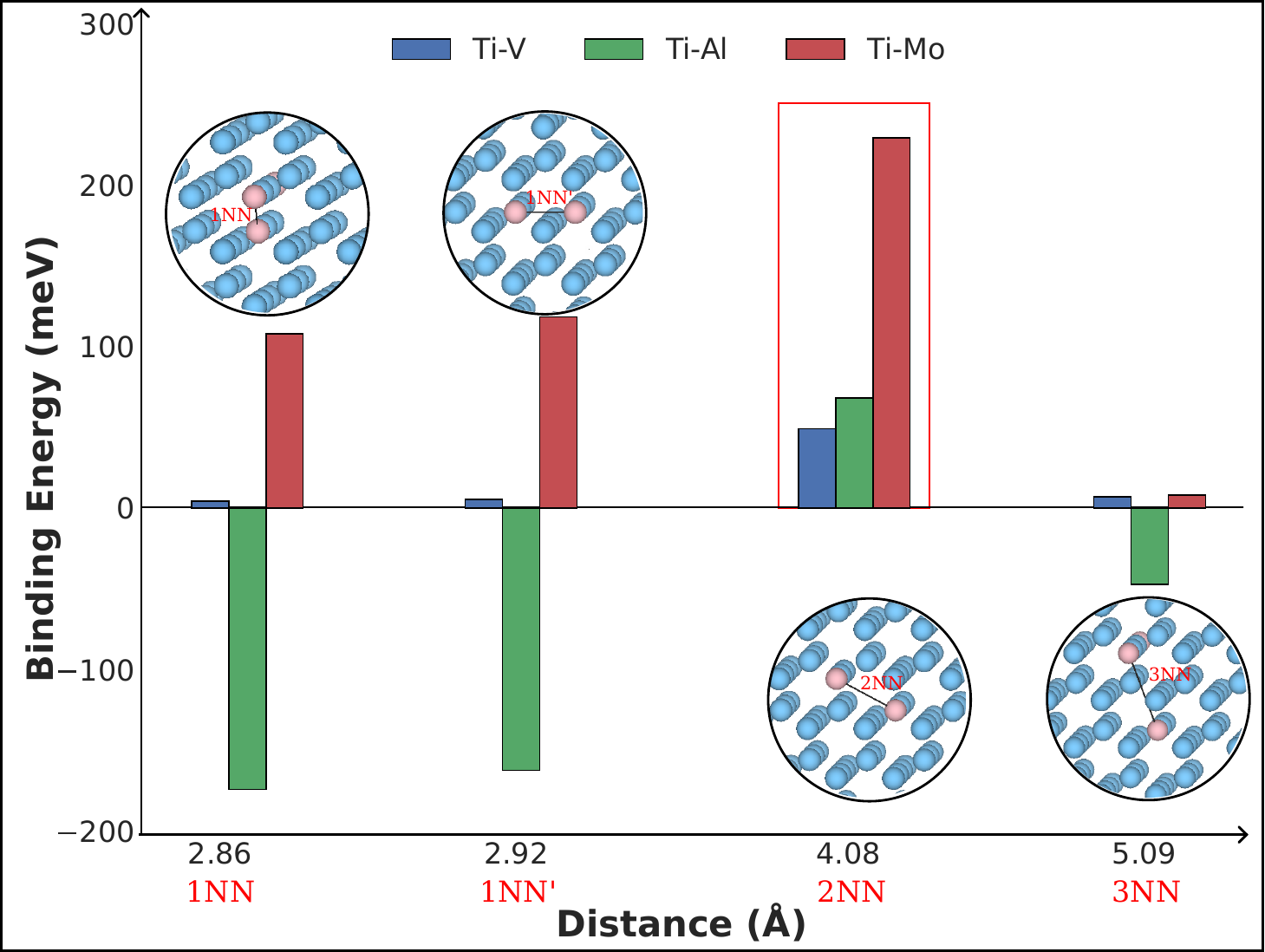}
\caption{Calculated binding energies ($E_b$) for binary Ti-X systems (X = Al, V, Mo) as a function of solute separation distance from 1NN to 3NN. Positive values indicate an attractive interaction, indicating a clustering tendency. Note the strong preference for 2$^{nd}$ nearest-neighbor (2NN) ordering for all substitutional solutes.}
\label{fig:binary_binding}
\end{figure}

Instead of forming a random solid solution, solute atoms sometimes prefer to form short-range order due to electronic structure effects and/or lattice-strain minimization. Such an ordering can have a profound effect on the mechanical properties. A systematic analysis of binding energy [Eq.~\ref{eq:Eb_binary}] reveals that Al atoms in $\alpha$-Ti have a strong preference for 2NN separation, while 1NN, 1NN', and 3NN are clearly unfavorable, with negative binding energy [Fig.~\ref{fig:binary_binding} and Table~\ref{tab:binding_energies_transposed}]. The calculated binding-energy difference between the favorable (2NN) and unfavorable configurations exceeds 150 meV/atom in Ti-Al, indicating a substantial thermodynamic driving force for SRO. The binding energy beyond 3NN is also calculated for the Ti-X system and is presented in Fig.~S1 of the SM, clearly showing 2NN as a favorable SRO. 

Mo and V substitutions in $\alpha$-Ti also exhibit SRO tendencies. Unlike Al, while Mo and V show positive binding energy at 1NN and 1NN', the preference from 2NN separation is evident from Fig.~\ref{fig:binary_binding} and Table~\ref{tab:binding_energies_transposed}. Our findings agree well with previous observations for Ti-Al systems~\citep{kwasniak2023competition}, while providing new insights for Ti-V and Ti-Mo systems that have received comparatively less attention in the literature. Such atomic-level details go beyond EET-based approaches that neglect SRO effects and assume random solid solutions, underscoring the critical importance of incorporating realistic atomic configurations into predictive models. 

\begin{table}
\centering
\caption{Binding energies ($E_b$) for Ti-X systems as a function of solute separation. Bold values indicate the most favorable location.}
\label{tab:binding_energies_transposed}
\setlength{\tabcolsep}{4pt} 
\renewcommand{\arraystretch}{1.3} 
\small
\begin{tabular}{@{}lc*{7}{r}@{}}
\toprule
\textbf{System} & \textbf{Distance (Å)} & \textbf{1NN (2.86)} & \textbf{1NN' (2.92)} & \textbf{2NN (4.08)} & \textbf{3NN (5.09)} & \textbf{4NN (5.47)} & \textbf{5NN (6.51)} & \textbf{6NN (7.71)} \\
\midrule
\textbf{Ti-Al} 
& $E_b$ (meV) 
& -174.00
& -162.20 
& \textbf{68.40}
& -47.20 
& 1.30 
& 10.40 
& -5.00 \\
\midrule
\textbf{Ti-V} 
& $E_b$ (meV) 
& 4.40 
& 5.60 
& \textbf{49.10} 
& 7.20 
& -1.40 
& 8.30 
& 5.20 \\
\midrule
\textbf{Ti-Mo} 
& $E_b$ (meV) 
& 108.00 
& 118.60 
& \textbf{229.30} 
& 8.20 
& 128.70 
& 91.70 
& 59.20 \\
\bottomrule
\end{tabular}
\end{table}

\begin{table}
\centering
\caption{Binding energies of Ti-X-Y as a function of distance d$_i$ [Fig.~\ref{fig:schematic_sro}]. Bold values indicate the most favorable location.}
\label{tab:binding_energies_heatmap}
\setlength{\tabcolsep}{4pt}
\renewcommand{\arraystretch}{1.3}
\small
\begin{tabular}{@{}lcccccccccccc@{}}
\toprule
\multirow{2}{*}{\textbf{}} & \multicolumn{4}{c}{\textbf{Ti-2Al (meV)}} & \multicolumn{4}{c}{\textbf{Ti-2V (meV)}} & \multicolumn{4}{c}{\textbf{Ti-2Mo (meV)}} \\
\cmidrule(lr){2-5} \cmidrule(lr){6-9} \cmidrule(lr){10-13}
& \textbf{d1} & \textbf{d2} & \textbf{d3} & \textbf{d4} & \textbf{d1} & \textbf{d2} & \textbf{d3} & \textbf{d4} & \textbf{d1} & \textbf{d2} & \textbf{d3} & \textbf{d4} \\
\midrule
\textbf{H} & -365.5 & -43.6 & 41.3 & \textbf{50.6} 
           & 13.1 & 22.5 & 14.5 & \textbf{39.3} 
           & -141.2 & 46.9 & 68.6 & \textbf{84.8} \\
\textbf{C} & -1131.2 & -574.5 & -35.7 & \textbf{27.0}
           & \textbf{254.4} & 189.3 & -27.3 & 47.1 
           & -155.6 & -45.5 & -83.9 & \textbf{38.0} \\
\textbf{N} & -1496.1 & -743.2 & -54.1 & \textbf{14.0} 
           & 21.4 & \textbf{48.7} & -20.7 & 31.9 
           & -582.2 & -218.0 & -59.9 & \textbf{25.2} \\
\textbf{O} & -1381.8 & -751.4 & -46.5 & \textbf{25.0}
           & -212.1 & -89.1 & -20.5 & \textbf{16.8}
           & -607.3 & -375.8 & -21.9 & \textbf{277.2} \\
\bottomrule
\end{tabular}
\end{table}

The introduction of an interstitial atom Y produces complex and non-uniform effects on the SRO of Ti-X systems, with behavior strongly dependent on the specific elemental combinations. 
For example, in Ti-Al-O systems, binding energy calculations [Eq.~\ref{eq:Eb_ternary}] reveal a substantial energetic penalty (exceeding 300 meV) for oxygen occupying the d$_1$ interstitial site over the d$_4$ interstitial site [Fig.~\ref{fig:schematic_sro} and Table~\ref{tab:binding_energies_heatmap}]. Note that the d$_1$ interstitial site has a mixed Ti-Al coordination, while the d$_4$ interstitial site has a purely Ti coordination.  Thus, the oxygen atoms preferentially occupy sites with pure Ti coordination, thereby creating oxygen-rich regions that segregate from aluminum clusters. While this result is consistent with previous findings~\citep{bakulin2022impurity, gunda2020understanding, bakulin2020diffusion}, we present a detailed mechanistic understanding through electronic structure analysis later in this paper. A very similar trend is observed for all Ti-Al-Y and Ti-Mo-Y systems, with the Y atom preferring to occupy the d$_4$ interstitial site [Fig.~\ref{fig:schematic_sro} and Table~\ref{tab:binding_energies_heatmap}]. The bar chart of the binding energy for the Ti-X-Y system from  d$_1$ to  d$_4$ is shown in Fig.~S2 of the SM. 

A notable exception to this general trend is observed in Ti-V-C systems. DFT binding-energy analysis reveals strong attractive interactions between carbon and the mixed Ti-V octahedral site (d$_1$), promoting distinct ternary SRO formation. Thus, carbon atoms in this system act as ordering centers that enhance and stabilize local Ti-V SRO, creating complex ternary ordered structures whose influence on mechanical response is mirrored in recent first-principles studies of ordered Ti$_{1-x}$V$_{x}$C phases~\citep{prysyazhnyuk2025first, wang2022elastic}, experimental observations of pronounced carbon-sublattice ordering and superlattice formation in (Ti, V)C$_{x}$ phases~\citep{burvsik1999ordering, bandyopadhyay2000ti}, and with theoretical treatments of interstitial-driven short-range ordering in transition-metal carbides~\citep{gusev1989short}. On the other hand, N prefers the d$_2$ site, while H and O prefer the d$_4$ site in Ti-V [Fig.~\ref{fig:schematic_sro} and Table~\ref{tab:binding_energies_heatmap}], making it the most non-trivial system in terms of interstitial occupation.

\subsection{Bonding Analysis and GSFE in Binary Ti-X and Ti-Y}
\begin{figure}
\centering
\includegraphics[width=\linewidth]{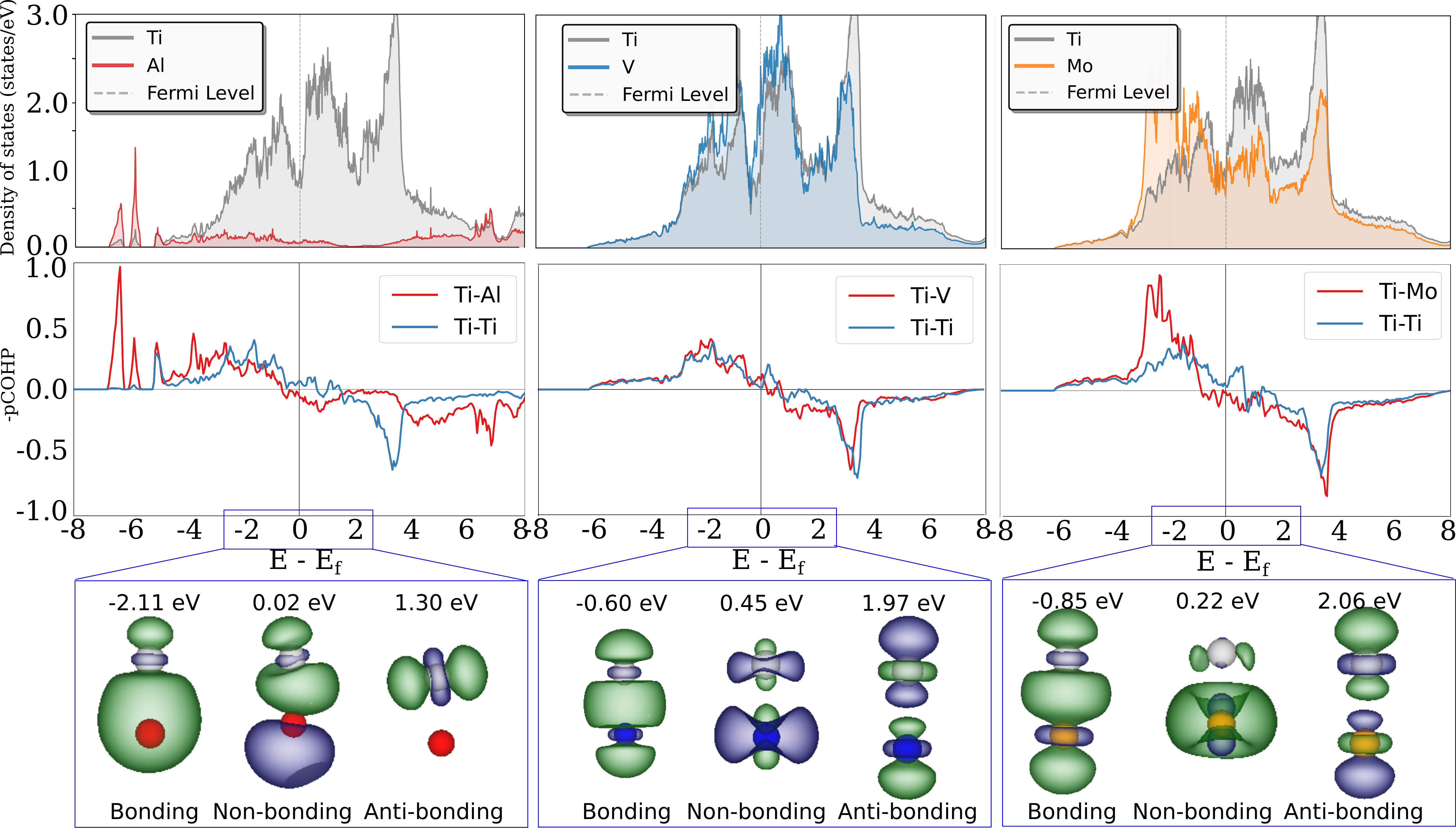}
\caption{Binary Ti-X systems. (Top) Projected density of states (pDOS) showing the contributions from Ti and X. (Middle) Projected Crystal Orbital Hamilton Population ($-p$COHP) curves illustrating bonding (positive values) and anti-bonding (negative values) interactions with respect to the Fermi level; red and black lines correspond to Ti-X and Ti-Ti bonds, respectively. The area under the $-p$COHP curve integrated up to the Fermi level gives the ICOHP value, which quantifies the overall bond strength: Ti-Mo (-1.288 eV) > Ti-Al (-1.101 eV) > Ti-V (-0.850 eV). (Bottom) Calculated orbitals (isosurface value = 0.08 e/\AA$^{3}$), showing bonding, non-bonding, anti-bonding interactions. Note the similarity between Ti-Mo and Ti-V orbitals.}
\label{fig:TiX_dos_cohp_mo}
\end{figure}

\begin{figure}
\centering
\includegraphics[width=\linewidth]{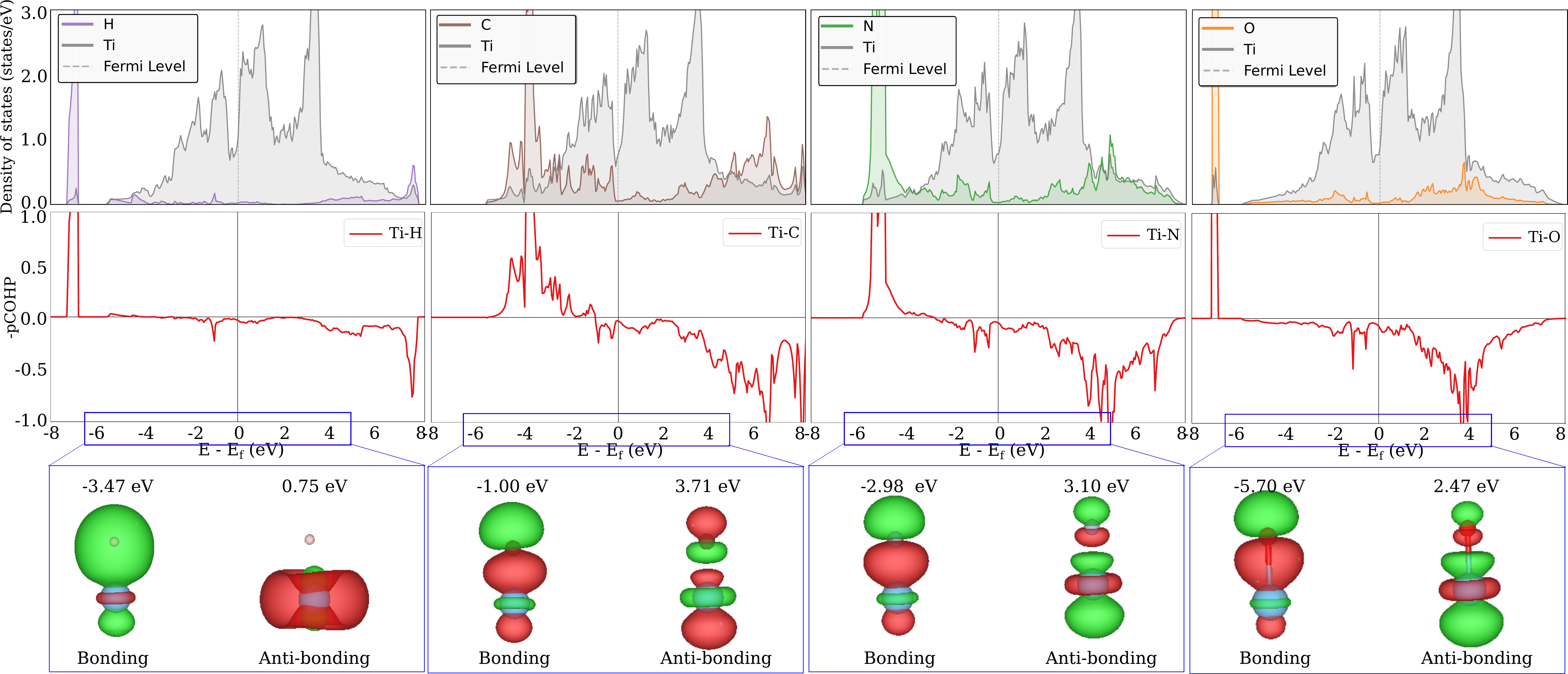}
\caption{Ti-Y systems. (Top) Projected density of states (pDOS) showing contributions from Ti and the Y element. (Middle) Projected Crystal Orbital Hamilton Population ($-p$COHP) curves illustrating bonding and anti-bonding character for Ti-Y interactions. C, N, and O interstitials show substantial bonding regions with ICOHP values: Ti-C (-2.58 eV) > Ti-N (-2.55 eV) > Ti-O (-2.035 eV). In contrast, H interstitial shows weak bonding character (ICOHP -0.707 eV). (Bottom) Calculated orbitals (isosurface value = 0.1 e/\AA$^{3}$), clearly showing the similarity among C, N, O, with strong directional p-d hybridization and concentrated electron density between Ti and interstitial atoms.}
\label{fig:Ti_Y_cohp_dos_MO}
\end{figure}

After studying the SRO, we now focus on the nature of Ti-X and Ti-Y bonding in terms of electronic structure to help us understand the GSFE trend observed in these systems. Binary Ti-Al, Ti-V, and Ti-Mo systems exhibit significantly different electronic bonding characteristics [Fig.~\ref{fig:TiX_dos_cohp_mo}]. \textcolor{black}{Ti-Mo displays the deepest bonding region in the $-$pCOHP curves, with a \textcolor{black}{binary} Ti-Mo ICOHP of \textcolor{black}{$-1.288$} eV compared to $-1.10$ eV for Ti-Al and $-0.86$ eV for Ti-V. This indicates a \textcolor{black}{$\sim17\%$ and $\sim52\%$} increase in bond strength of Ti-Mo over Ti-Al and Ti-V, respectively}. In binary Ti-Y systems [Fig.~\ref{fig:Ti_Y_cohp_dos_MO}], Ti-interstitial bonding strength follows the hierarchy C $\approx$ N $>$ O $\gg$ H, as reflected in ICOHP values of $-2.58$ eV (Ti-C), $-2.55$ eV (Ti-N), $-2.03$ eV (Ti-O), and only $-0.71$ eV (Ti-H). The rest of the section focuses on how Ti-X and Ti-Y ICOHP values dictate the GSFE curves calculated for these systems. 

\begin{figure}
\centering
\includegraphics[width=0.75\linewidth]{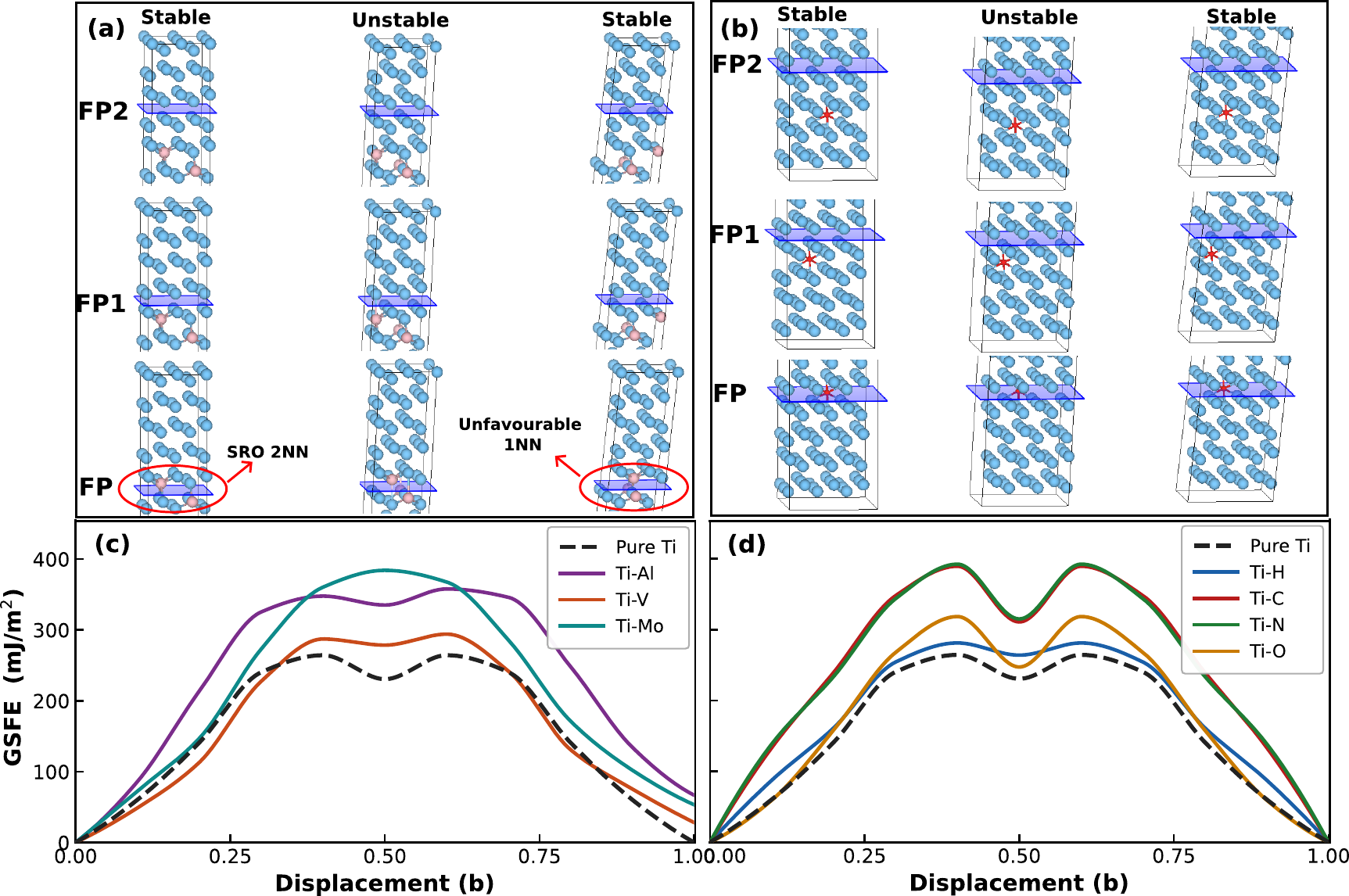}
\caption{Evolution of bonding during slip and GSFE for $\alpha$-Ti alloys. (a)~Atomic configurations of substitutional Ti-X (X~=~Al, V, Mo), shown at the initial, unstable (mid-slip), and final states as the upper portion is displaced by one Burgers vector. FP, FP1, and FP2 are three different locations of the two X atoms (red circles) relative to the slip plane.  In the initial configuration, the two substitutional solute atoms reside at their energetically preferred 2NN SRO positions, which are disrupted by the slip process. (b)~Atomic configurations for interstitial Ti-Y (Y~=~H, C, N, O). Initially, the interstitial atom (red cross) occupies an octahedral site, transitions through an unstable state, and finally returns to an equivalent octahedral site. Prismatic GSFE curves for (c) Ti-X, and (d) Ti-Y, plotted alongside pure $\alpha$-Ti (black dashed).}
\label{fig:gsfe_TiXY}
\end{figure}

Fig.~\ref{fig:gsfe_TiXY} (a) and (c) show the prismatic GSFE curves for pure $\alpha$-Ti and Ti-X alloys (X~=~Al, V, Mo), along with a schematic illustrating the evolution of the local bonding environment as the upper half-crystal is rigidly displaced by one Burgers vector. The energy barrier is calculated using Eq.~\ref{eq:gsfe}. As shown in Fig.~\ref{fig:gsfe_TiXY} (a), we consider three different configurations: FP (both the solute atoms X, located on the slip plane), FP1 (one of the solute atoms X, located on the slip plane), and FP2 (none of the solute atoms X, located on the slip plane). We evaluate the average GSFE across all three fault-plane positions, $\bar{\gamma}_{\mathrm{usf}} = \frac{1}{3}\left[\gamma_{\mathrm{usf}}^{\mathrm{FP}} + \gamma_{\mathrm{usf}}^{\mathrm{FP1}} + \gamma_{\mathrm{usf}}^{\mathrm{FP2}}\right]$, representing a better statistics of slip resistance in such systems.

As shown in Fig.~\ref{fig:gsfe_TiXY} (c), the energy barrier for slip to take place is the highest for Ti-Mo, followed by Ti-Al and Ti-V. Quantitatively, the USFE of pure Ti for prismatic slip is 264~mJ/m$^2$, which increases to 360~mJ/m$^{2}$ (Ti-Mo), 347~mJ/m$^{2}$ (Ti-Al), and 287~mJ/m$^{2}$ (Ti-V) in different binary systems. The USFE hierarchy is consistently captured by calculated bond strength $\propto$ |ICOHP|: Ti-Mo~(\textcolor{black}{1.288}~eV)~>~Ti-Al~(1.10~eV)~>~Ti-V~(0.86~eV), demonstrating the intrinsic connection between electronic bonding and slip barriers. This one-to-one correspondence between the slip energy barrier and the magnitude of ICOHP establishes the latter as a physically meaningful and reliable descriptor of solid-solution strengthening in $\alpha$-Ti. 

Interestingly, SRO also affects the GSFE curves. As shown in Fig.~\ref{fig:gsfe_TiXY} (c), the initial and final energies for pure Ti are the same, as both the structures are identical after moving the upper portion by one Burgers vector. This is not the case for Ti-X alloys, where the energy is higher (compared to the initial structure) after displacing the upper portion by one Burgers vector. Note that in the initial configuration, the two substitutional atoms reside at their energetically preferred SRO positions, separated by a 2NN distance. As a result of the slip, SRO is disrupted, leading to an atomic arrangement having higher energy than the initial configuration. As reported in Table~\ref{tab:binding_energies_transposed}, the Ti-Al system has the highest energy penalty for breaking SRO, followed by Ti-Mo and Ti-V. Accordingly, energy after slip by one Burgers vector is the highest for Ti-Al, followed by Ti-Mo and Ti-V [Fig.~\ref{fig:gsfe_TiXY} (c)].

Fig.~\ref{fig:gsfe_TiXY} (b) and (d) show the prismatic GSFE curves for pure $\alpha$-Ti and Ti-Y systems (Y~=~H, C, N, O, located in the octahedral void) and evolution of the local bonding environment as the upper half-crystal is rigidly displaced by one Burgers vector. The energy barrier is calculated using Eq.~\ref{eq:gsfe}. Similar to the previous case, we consider three different configurations: starting with FP, when Y is located on the slip plane, then moving it systematically away from the slip plane at FP1 and FP2 [Fig.~\ref{fig:gsfe_TiXY} (b)], and finally report the average $\bar{\gamma}_{\mathrm{usf}}$. 

The GSFE curves for the Ti-Y systems are shown in Fig.~\ref{fig:gsfe_TiXY}(d). In contrast to the substitutional Ti-X alloys, the Ti-Y GSFE curves are symmetric about the midpoint of the Burgers vector. This symmetry arises because no SRO is involved with interstitial atoms. As a result of the atomic movements involved with the slip process, the interstitial atom simply migrates from its initial octahedral site through a transient state and back into an equivalent octahedral site, a path that produces a symmetric energy profile. Despite this structural difference, the average USFE of Ti-Y systems follows the same ICOHP-governed trend as Ti-X. The interstitial strengthening hierarchy is Ti-C~$\approx$~Ti-N~$>$~Ti-O~>~Ti-H [Fig.~\ref{fig:gsfe_TiXY}(d)], directly mirroring the ICOHP values: $-2.58$ (Ti-C), $-2.55$ (Ti-N), $-2.03$ (Ti-O), and $-0.71$~eV (Ti-H). 

We have also calculated the GSFE curves for the basal slip system (refer to Fig. S3 of the SM), which is the secondary slip system in $\alpha$-Ti. We have observed a similar trend for Ti-X and Ti-Y basal slip systems: higher |ICOHP| leading to higher USFE. This further consolidates ICOHP as a reliable descriptor for strengthening in both substitutional and interstitial $\alpha$-Ti alloys and justifies its use as the central parameter in our empirical strength model, to be presented in a later section.

\subsection{Bonding Analysis in Ternary Ti-X-Y}


\begin{table}
\centering
\caption{Integrated Crystal Orbital Hamilton Population (ICOHP) values (eV) for Ti-X-Y ternary systems in two configurations: (i) interstitial Y located at the Ti-X coordinated octahedral site, and (ii) interstitial Y located away from the Ti-X octahedral site. Ti-X denotes Ti-Al, Ti-V, and Ti-Mo bonds; Ti-Y denotes Ti-interstitial bonds (H, C, N, O). More negative values indicate stronger bonding interactions.}
\label{tab:icohp_ternary_manual}
\setlength{\tabcolsep}{4pt}
\renewcommand{\arraystretch}{1.2}
\small

\begin{tabular}{@{}lcccccccccccc@{}}
\toprule
\multirow{2}{*}{\textbf{Bond Type}} & \multicolumn{4}{c}{\textbf{Ti-Al-Y}} & \multicolumn{4}{c}{\textbf{Ti-V-Y}} & \multicolumn{4}{c}{\textbf{Ti-Mo-Y}} \\
\cmidrule(lr){2-5} \cmidrule(lr){6-9} \cmidrule(lr){10-13}
& H & C & N & O & H & C & N & O & H & C & N & O \\
\midrule

\multicolumn{13}{c}{\textbf{Y at Ti-X octa}} \\
\textbf{Ti-X}$^{a}$ 
& $-1.127$ & $-1.102$ & $-1.107$ & $-1.160$ 
& $-0.938$ & $-0.856$ & $-0.855$ & $-0.876$ 
& $-1.353$ & $-1.432$ & $-1.544$ & $-1.612$ \\

\textbf{Ti-Y}$^{b}$ 
& $-0.659$ & $-2.424$ & $-2.358$ & $-1.861$ 
& $-0.778$ & $-2.493$ & $-2.537$ & $-2.114$ 
& $-0.778$ & $-2.503$ & $-2.730$ & $-2.341$ \\
\midrule

\multicolumn{13}{c}{\textbf{Y away from Ti-X octa}} \\
\textbf{Ti-X}$^{a}$ 
& $-1.092$ & $-1.125$ & $-1.117$ & $-1.112$ 
& $-1.100$ & $-1.115$ & $-1.109$ & $-1.106$ 
& $-1.599$ & $-1.578$ & $-1.587$ & $-1.581$ \\

\textbf{Ti-Y}$^{b}$ 
& $-0.769$ & $-2.822$ & $-2.731$ & $-2.135$ 
& $-0.725$ & $-2.765$ & $-2.707$ & $-2.125$ 
& $-0.723$ & $-2.775$ & $-2.718$ & $-2.134$ \\
\bottomrule
\end{tabular}
\vspace{0.05cm}
\footnotesize
\begin{tabular}{@{}p{16cm}@{}}
\toprule
Ti-Ti bond interaction: Ti-Ti = $-0.76$\\
$^{a}$ Denotes Ti-X (substitutional) bond interactions. Binary reference: Ti-Al = $-1.101$, Ti-V = $-0.850$, Ti-Mo = $-1.288$. \\
$^{b}$ Denotes Ti-Y (interstitial) bond interactions. Binary reference: Ti-H = $-0.707$, Ti-C = $-2.587$, Ti-N = $-2.555$, Ti-O = $-2.035$. \\

\bottomrule
\end{tabular}
\end{table}



The ternary Ti-X-Y systems introduce an additional layer of complexity because the interstitial can occupy octahedral sites either near or away from the Ti-X SRO cluster. Binding energy (Table~\ref{tab:binding_energies_heatmap}) determines the thermodynamically preferred position: for Ti-Al-Y and Ti-Mo-Y (Y = H, O, C, N), the interstitial experiences repulsion from the Ti-X 2NN cluster and prefers to sit at the Ti-rich octahedral site $d_{4}$ [Fig.~\ref{fig:binary_binding}]. In the case of Ti-V-Y, we see some exceptions. For example, carbon prefers the Ti-V coordinated octahedron $d_{1}$, nitrogen prefers $d_2$, while H and O prefer $d_4$. 

These thermodynamically stable geometries directly influence ICOHP. For example, in Ti-Al-Y, the Ti-Al and Ti-Y ICOHPs depend on whether the Y atom occupies the $d_1$ (non-preferred) or $d_4$ (preferred) site. Comparing with the binary reference values, we find that the variation in Ti-Al ICOHP is less sensitive to the location of Y than that of the Ti-Y ICOHP [Table~\ref{tab:icohp_ternary_manual}]. For example, depending on whether Y occupies the $d_1$ or $d_4$ position, the Ti-Y |ICOHP| values decrease or increase up to 6-9\% compared to binary reference values. In contrast, only up to a 2-5\% increase in magnitude is observed for Ti-Al |ICOHP| (with respect to the binary reference of $-1.101$ eV) when Y occupies the $d_4$ or $d_1$ position. Thus, ICOHP quantitatively reflects the geometric and electronic changes imposed by the interstitial’s location.


Interestingly, a different trend emerges in Ti-V-Y. Depending on whether Y occupies the $d_1$ or $d_4$ position, the Ti-Y |ICOHP| values change by 1-8\% (plus or minus) compared to binary reference values. These numbers are comparable to the Ti-Al-Y system. However, a very sharp increase ($\sim 30\%$) of magnitude is observed for Ti-V |ICOHP| (with respect to the binary reference of $-0.85$ eV) when Y occupies the $d_4$ position, while a much smaller change is observed when Y occupies the $d_1$ position. This trend is unlike Ti-Al-Y, where Ti-Al ICOHP shows only a mild effect on the location of Y. 



In the case of Ti-Mo-Y, depending on whether Y occupies the $d_1$ or $d_4$ position, the Ti-Y |ICOHP| values change by 3-10\% (plus or minus) compared to binary reference values. These numbers are comparable to the Ti-Al-Y and Ti-V-Y systems. Similar to Ti-V-Y, a very sharp increase ($\sim 25\%$) of magnitude is observed for Ti-Mo |ICOHP| (with respect to the binary reference of $-1.288$ eV) when Y occupies the $d_4$ position. However, unlike Ti-V-Y, Ti-Mo |ICOHP| also increases by $\sim 26\%$ when Y occupies the $d_1$ position, particularly for Y = N and O. Such non-trivial behavior across different X and Y underscores the necessity of first-principles calculations to capture the details of electronic interactions in these systems.

\textcolor{black}{To characterize the bonding scenario without relying on a single descriptor, Mulliken and L\"owdin population analyses were carried out for all twelve ternary systems; the complete set of atomic charges is listed in Table~S3 of the SM. Read together with the bond indices of Table~\ref{tab:icobi_values}, the two descriptors resolve three distinct bonding regimes among the substitutional solutes. Aluminum carries by far the largest charge ($-1.68$ Mulliken, $-1.58$ L\"owdin) while exhibiting the smallest bond index (ICOBI $\approx 0.26$-$0.28$), identifying Ti-Al as the most polar and least covalent of the three substitutional interactions. Molybdenum combines a substantial charge transfer ($-1.00$, $-0.94$) with the highest bond index (ICOBI $\approx 0.52$-$0.63$), the signature of a strongly polar covalent bond, which is consistent with its deep $d$-$d$ bonding region and the most negative Ti-X ICOHP. Vanadium is almost electroneutral ($-0.13$, $-0.11$) at an intermediate bond index (ICOBI $\approx 0.39$-$0.43$), corresponding to an essentially non-polar, metallic Ti-V interaction.}

\textcolor{black}{For the interstitials, the charges follow the sequence C ($-1.35$) $>$ N ($-1.17$) $>$ O ($-0.97$) $\gg$ H ($-0.48$). This ordering is the reverse of the electronegativity sequence O $>$ N $>$ C, which rules out a simple ionic description and confirms that the Ti-Y interaction is governed by directional $p$-$d$ covalency rather than by charge transfer. The charges instead track the ICOHP and ICOBI hierarchies, both of which also decrease along C $\rightarrow$ N $\rightarrow$ O. This concurrent decrease has a transparent electronic origin: progressing from C to N to O adds valence $p$ electrons to the system, and these additional electrons progressively populate Ti-Y antibonding states lying below the Fermi level. The extra electrons, therefore, do not reinforce the bond but partially cancel it, so that the shared-electron count (ICOBI) and the integrated bond strength (ICOHP) fall even as the nominal electronegativity of the interstitial rises.}


\begin{table}
\centering
\caption{Integrated Crystal Orbital Bond Index (ICOBI) values for Ti-X-Y ternary systems showing bond indices in different bond environments.}
\label{tab:icobi_values}
\setlength{\tabcolsep}{5pt}
\renewcommand{\arraystretch}{1.3}
\small
\begin{tabular}{@{}lcccccccccccc@{}}
\toprule
\multirow{2}{*}{\textbf{Bond Type}} & \multicolumn{4}{c}{\textbf{Ti-Al-Y}} & \multicolumn{4}{c}{\textbf{Ti-V-Y}} & \multicolumn{4}{c}{\textbf{Ti-Mo-Y}} \\
\cmidrule(lr){2-5} \cmidrule(lr){6-9} \cmidrule(lr){10-13}
& \textbf{H} & \textbf{C} & \textbf{N} & \textbf{O} & \textbf{H} & \textbf{C} & \textbf{N} & \textbf{O} & \textbf{H} & \textbf{C} & \textbf{N} & \textbf{O} \\
\midrule
\textbf{Ti-Ti} & 0.365 & 0.381 & 0.383 & 0.394
               & 0.353 & 0.333 & 0.351 & 0.357 
               & 0.404 & 0.364 & 0.376 & 0.400 \\
\textbf{Ti-X}  & 0.271 & 0.269 & 0.261 & 0.281
               & 0.431 & 0.394 & 0.395 & 0.399 
               & 0.515 & 0.551 & 0.601 & 0.629 \\
\textbf{Ti-Y}  & 0.121 & 0.523 & 0.442 & 0.290 
               & 0.128 & 0.170 & 0.422 & 0.300 
               & 0.130 & 0.478 & 0.472 & 0.343 \\
\midrule
\multicolumn{13}{l}{\small \textit{Note: Ti-X refers to Ti-Al, Ti-V, or Ti-Mo bonds. Ti-Y refers to Ti-interstitial bonds (Ti-H, Ti-C, Ti-N, Ti-O).}} \\
\bottomrule
\end{tabular}
\end{table}

\begin{figure}
\centering
\includegraphics[width=\linewidth]{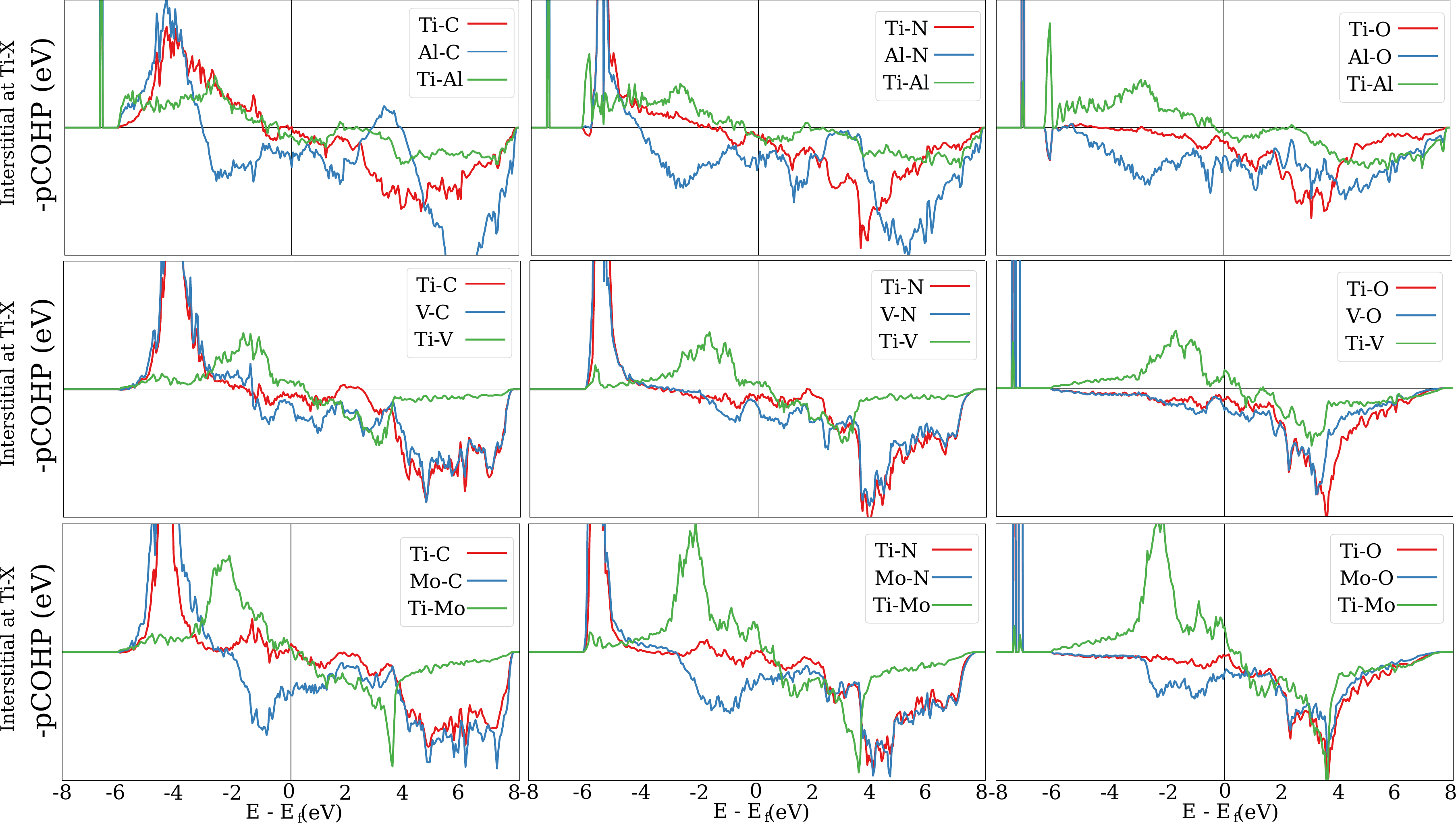}
\caption{The $-p$COHP plots of d$_{1}$ configuration for (1) Ti-Al-C/N/O, (2) Ti-V-C/N/O, and (3) Ti-Mo-C/N/O.  }
\label{fig:[pcohp_plot]}
\end{figure}

Interestingly, $-p$COHP curves also show an electronic signature of attractive/repulsive interaction between the atomic pairs in ternary systems. For example, in the case of Ti-Al-O, which is energetically unfavorable in $d_1$ configuration [$E_b = -1382$~meV, Table~\ref{tab:binding_energies_heatmap}], the $-p$COHP curves reveal a well-developed Ti-Al bonding region, but a pronounced Al-O \textit{anti-bonding} region appears simultaneously [Fig.~\ref{fig:[pcohp_plot]}]. Such a trend directly reflects the electronic incompatibility between O~$2p$ and Al~$3p$ states in $d_1$ configuration. This anti-bonding character serves as the electronic fingerprint of the repulsive Al-O interaction in the ternary Ti-Al-O system. 
In stark contrast, the $d_1$ configuration is energetically favorable in the case of Ti-V-C [$E_b = +254$~meV], and V-C \textit{bonding} regions are clearly visible in the $-p$COHP curve, which serves as the electronic fingerprint of the attractive V-C interaction in the ternary Ti-V-C system. 

\subsection{Electronic Origin of Strengthening}
Experimental evidence consistently shows that alloying elements can significantly improve the tensile strength of titanium and alter its plastic response characteristics. In this section, we argue, from the point of view of electronic structure, that substitutional (X) and interstitial (Y) atoms in the $\alpha$-Ti lattice lead to strengthening, mainly due to an increase in energy barrier ($\Delta E$) for dislocation core movement. This energy barrier primarily comprises two components: resistance to dislocation nucleation and resistance to dislocation gliding.


\begin{figure}
    \centering
    \includegraphics[width=0.8\linewidth]{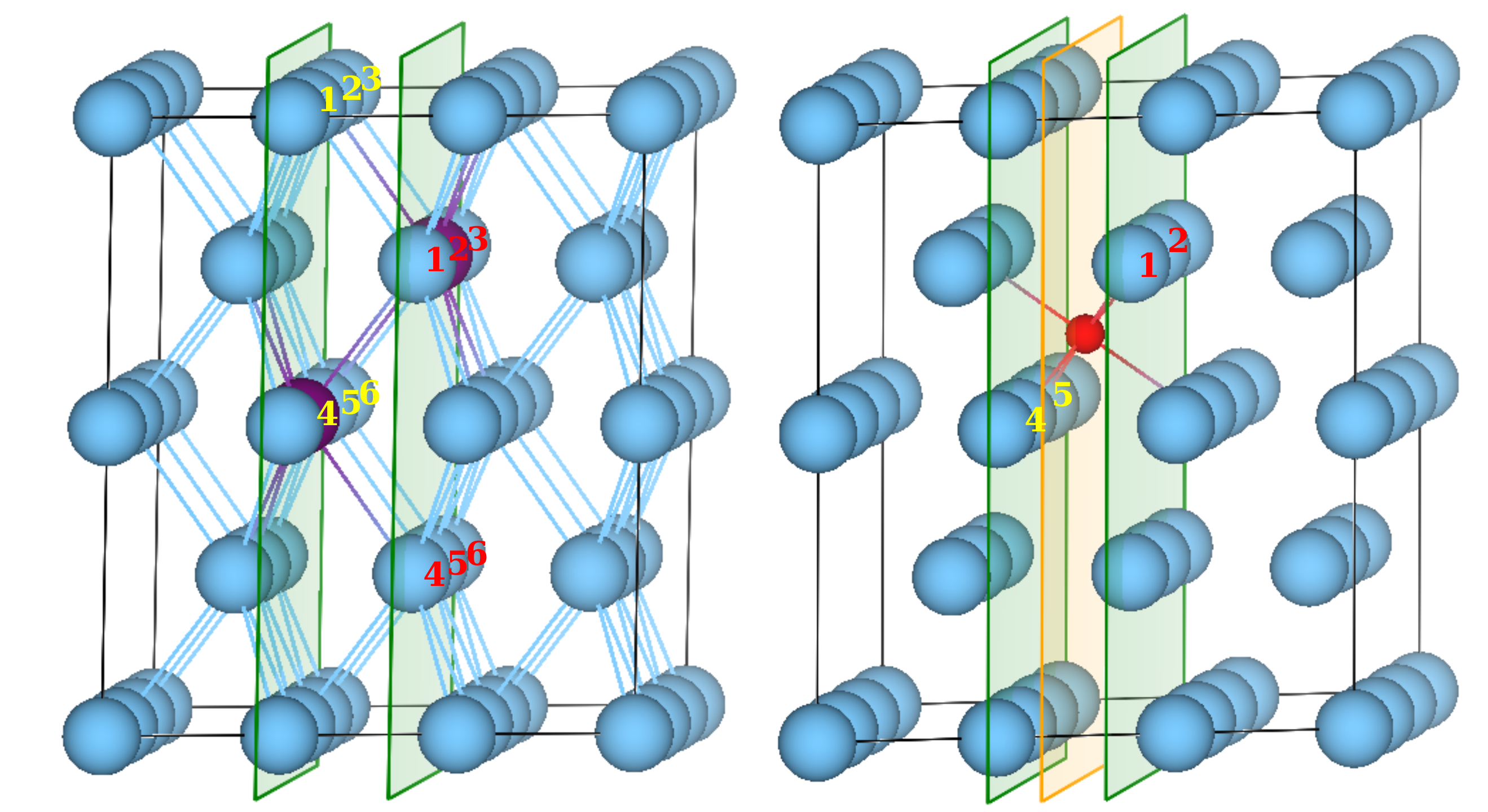}
    \caption{Schematic illustration of the ``interfacial bonds'' used to evaluate the planar ICOHP density $\eta^{(hkil)}$ and the interfacial coefficient $\Delta\eta$. Left: Two adjacent wide prismatic planes (green), containing Ti and substitutional X atoms. Atoms 1-3 (red) each form one bond with atoms 1-3 (yellow, upper) and one bond with atoms 4-6 (yellow, lower), producing Ti-Al and Ti-Ti bonds in the ``interfacial region''. The planar ICOHP density $\eta^{(hkil)}$ is obtained by summing ICOHP contributions of these nearest-neighbor bonds per unit area (Eq.~\ref{eq:eta_icohp}), and its normalized difference relative to the pure Ti defines $\Delta\eta_{\mathrm{Ti-X}}$ (Eq.~\ref{eq:delta_eta_icohp}). Right: An intermediate plane (orange) between the two adjacent wide prismatic planes hosts the interstitial Y atom, with additional Ti-Y bonds in the ``interfacial region''. Including these additional Ti-Y ICOHP contributions defines $\eta^{(hkil)}$ (Eq.~\ref{eq:eta_icohp}) and comparing against the interstitial-free environment defines $\Delta\eta_{\mathrm{Ti-Y}}$ (Eq.~\ref{eq:delta_eta_icohp}).}
    \label{fig:eta}
\end{figure}

To evaluate the effects of alloying elements on the atomic-bond configuration and electronic structure of the $\alpha$-Ti lattice, this work moves beyond previous semi-empirical frameworks, such as valence-electron-parameter-based Empirical Electron Theory (EET). Instead, we use DFT-based electronic parameters (ICOHP values), which provide a direct, quantitative measure of bond strength derived from quantum mechanical electronic structure. By comparing ICOHP values of ``alloyed'' $\alpha$-Ti-X|Y regions with those of pure $\alpha$-Ti regions, strengthening mechanisms can be clearly revealed and quantified at a fundamental level. 

We construct a quantitative tensile strength model based on the idea that the critical stress for dislocation nucleation or glide is related to the energy required to break the strongest bonds in the material. We define two key descriptors derived from ICOHP: the intrinsic bond-strength coefficient $\gamma_{X|Y}$, which measures the relative strengthening of Ti-X or Ti-Y with respect to pure Ti, and the interfacial coefficient $\Delta \eta_{X|Y}$, which takes into account the change in bond density on preferred slip planes.

Specifically, let $|\text{ICOHP}|_{Ti-X|Y}^{}$ be the magnitude of the most negative ICOHP in the alloyed Ti lattice, and $|\text{ICOHP}|_{Ti}^{}$ be the corresponding value in pure $\alpha$-Ti. We define a dimensionless factor,
\begin{equation}
    \gamma_{Ti-X|Y} = \left[\frac{|\text{ICOHP|}_{Ti-X|Y}^{}}{|\text{ICOHP|}_{Ti}^{}} - 1\right ]
\label{eq:gamma_icohp}
\end{equation}
which represents the increase in bond strength of the strongest bonds due to the solute(s) and plays a critical role in dislocation nucleation or glide. 


An ``interfacial'' stress coefficient ($\Delta \eta_{X|Y}$) is used to evaluate the change in lattice resistance to dislocation gliding due to the substitutional/interstitial atom. This coefficient is calculated from the difference in the planar ICOHP density across the slip planes of the adjacent regions. The ICOHP density, $\eta_{\text{}}^{(hkil)}$, which represents the total first nearest neighbor bond strength per unit area of a slip plane, is defined as:
\begin{equation}
    \eta_{\text{}}^{(hkil)} = \frac{\sum_a |\text{ICOHP}_a| \times I_a}{S^{(hkil)}}
    \label{eq:eta_icohp}
\end{equation}
where $|\text{ICOHP}_a|$ is the absolute ICOHP value of the first nearest neighbor bond in the slip plane, $I_a$ is the number of such bonds, and $S^{(hkil)}$ is the geometric area of the selected slip plane (hkil).


Fig.~\ref{fig:eta} provides a schematic representation of how the bonds are partitioned when evaluating the interfacial coefficient. For substitutional alloys (Ti-X), the solute perturbs the Ti-Ti bonding network over a broader coordination shell, and only a subset of these bonds intersect the slip or ``interface'' plane. In contrast, interstitial solutes (Ti-Y) create a much more localized distortion field, producing a compact zone that interacts directly with the slip or ``interface'' plane. We only consider the prismatic slip system $\{10\bar{1}0\}\langle 11\bar{2}0 \rangle$, which is the primary slip system of Ti. The interfacial stress coefficient ($\Delta \eta_{X|Y}$) between the alloyed and pure regions can then be calculated as follows:
\begin{equation}
    \Delta\eta_{\text{Ti-X|Y}} = \frac{|\eta_{\text{} \text{Ti-X|Y}}^{(hkil)} - \eta_{\text{} \text{Ti}}^{(hkil)}|}{\frac{1}{2} \times \left(\eta_{\text{} \text{Ti-X|Y}}^{(hkil)} + \eta_{\text{} \text{Ti}}^{(hkil)}\right)}.
    \label{eq:delta_eta_icohp}
\end{equation}
This parameter provides a quantitative measure of the resistance to dislocation glide arising from local variations in the electronic bonding environment.

The final empirical formula for the tensile strength increment is taken as a linear combination of these coefficients: 
\begin{equation}
  \boxed{\sigma_{\rm tensile} = \sigma_0 + W_\gamma\,\sigma_0\,\gamma_{\rm Ti-X|Y} + W_\eta\,\sigma_0\,\Delta\eta_{\rm Ti-X|Y},}  
\end{equation}
where $\sigma_0$ is the strength of pure $\alpha$-Ti, and $W_\gamma$, $W_\eta$ are fitting weights that account for solute concentration factors. The key point is that each term now has a clear physical meaning tied to bond energies rather than an abstract empirical parameter.

\begin{figure}
    \centering
    \includegraphics[width=1\linewidth]{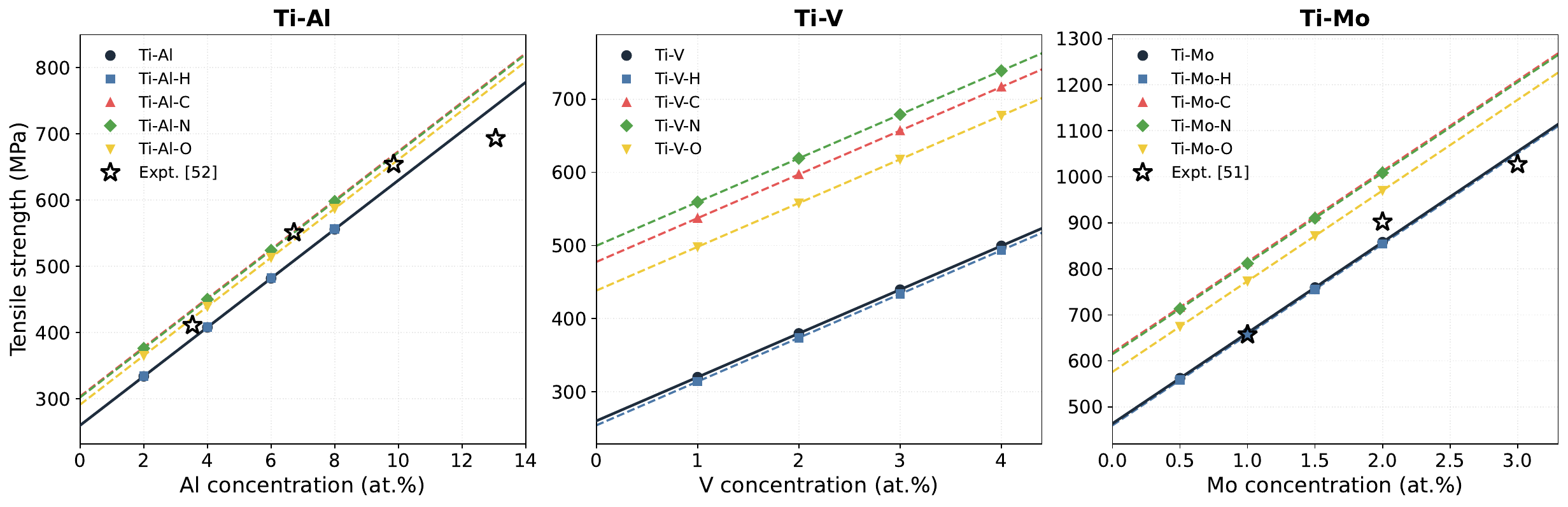}
    \caption{\textcolor{black}{Predicted tensile strength (solid line, substitutional binary; dashed lines, with a fixed 0.25~at.\% interstitial Y = H, C, N, O) as a function of substitutional alloying element (X) concentration, expressed throughout in atomic percent. Stars denote experimental data. Left: Ti-Al, compared with the binary Ti-Al series of Youssef \textit{et al.}~\citep{youssef2021effects}, whose compositions are reported in wt.\% and have been converted to at.\% using their measured Al contents (2.02, 3.90, 5.80, and 7.80~wt.\% Al, corresponding to 3.53, 6.72, 9.85, and 13.06~at.\% Al); mean deviation 6.1\%. Center: Ti-V, for which no binary tensile series is available and $p = q = 1$ is retained as an uncalibrated reference value, so that this panel is an upper bound rather than a validated prediction. Right: Ti-Mo, compared with Kobayashi and Okano~\citep{kobayashi2024effects}, whose compositions are already reported in at.\%; mean deviation 2.7\%. The Ti-Mo panel is referenced to the pure-Ti strength of 464~MPa measured in that same study, whereas the Ti-Al and Ti-V panels are referenced to $\sigma_0 = 260$~MPa~\citep{huang2023strengthening}. Within each host, the ordering of interstitial potency (Ti-C/N/O $\gg$ Ti-H) follows the ICOHP bonding strength hierarchy of Table~\ref{tab:icohp_ternary_manual}.}}
\label{fig:strength_predictions}
\end{figure}

Fig.~\ref{fig:strength_predictions} compares the computational strength predicted by our model with experimental results for a range of $\alpha$-type Ti-X-Y alloys from the literature. \textcolor{black}{All concentrations entering the model are atomic percentages; where the experimental compositions are reported in weight percent, they have been converted before comparison. Each series is referenced to the pure-Ti strength of the study supplying its validation data: $\sigma_0 = 260$~MPa for Ti-Al and Ti-V~\citep{huang2023strengthening}, and $\sigma_0 = 464$~MPa for Ti-Mo, the value measured by Kobayashi and Okano~\citep{kobayashi2024effects} for their own oxygen-bearing pure Ti. Thus, the model is compared against strength increments measured on the same material. The resulting mean deviations, 6.1\% for Ti-Al and 2.7\% for Ti-Mo, demonstrate that the underlying physics, such as binding-energy-driven site selection, ICOHP-controlled bond strength, and GSFE-governed slip resistance, form a coherent, quantitatively predictive framework for alloy design.}
\begin{table}
\centering
\caption{Tensile strength (MPa) of titanium alloys with substitutional and interstitial alloying elements. \textcolor{black}{All substitutional concentrations are atomic percentages. The Ti-Al and Ti-V columns are referenced to $\sigma_0 = 260$~MPa~\citep{huang2023strengthening}; the Ti-Mo column is referenced to $\sigma_0 = 464$~MPa, the pure-Ti strength measured in the same study that provides the Ti-Mo validation data~\citep{kobayashi2024effects}. The interstitial columns correspond to a fixed addition of 0.25~at.\% Y, evaluated with the same $p = q$ as the parent Ti-X system.}}
\label{tab:ti_alloys_comprehensive}
\resizebox{\textwidth}{!}{%
\begin{tabular}{@{}lccccc|lccccc|lccccc@{}}
\toprule
\multicolumn{6}{c}{\textbf{Ti-Al System}} & \multicolumn{6}{c}{\textbf{Ti-V System}} & \multicolumn{6}{c}{\textbf{Ti-Mo System}} \\
\cmidrule(lr){1-6} \cmidrule(lr){7-12} \cmidrule(lr){13-18}
\textbf{Alloy} & \textbf{Ti-Al} & \textbf{H} & \textbf{C} & \textbf{N} & \textbf{O} & 
\textbf{Alloy} & \textbf{Ti-V} & \textbf{H} & \textbf{C} & \textbf{N} & \textbf{O} & 
\textbf{Alloy} & \textbf{Ti-Mo} & \textbf{H} & \textbf{C} & \textbf{N} & \textbf{O} \\
\midrule
\textcolor{black}{Ti-2Al} & \textcolor{black}{333.9} & \textcolor{black}{334.2} & \textcolor{black}{377.8} & \textcolor{black}{376.2} & \textcolor{black}{365.2} & \textcolor{black}{Ti-1V} & \textcolor{black}{319.9} & \textcolor{black}{313.8} & \textcolor{black}{537.3} & \textcolor{black}{559.4} & \textcolor{black}{498.1} & \textcolor{black}{Ti-0.5Mo} & \textcolor{black}{562.5} & \textcolor{black}{558.5} & \textcolor{black}{716.8} & \textcolor{black}{713.1} & \textcolor{black}{674.7} \\
\textcolor{black}{Ti-4Al} & \textcolor{black}{407.9} & \textcolor{black}{408.2} & \textcolor{black}{451.7} & \textcolor{black}{450.1} & \textcolor{black}{439.2} & \textcolor{black}{Ti-2V} & \textcolor{black}{379.7} & \textcolor{black}{373.7} & \textcolor{black}{597.2} & \textcolor{black}{619.2} & \textcolor{black}{558.0} & \textcolor{black}{Ti-1.0Mo} & \textcolor{black}{661.0} & \textcolor{black}{657.0} & \textcolor{black}{815.3} & \textcolor{black}{811.6} & \textcolor{black}{773.2} \\
\textcolor{black}{Ti-6Al} & \textcolor{black}{481.8} & \textcolor{black}{482.1} & \textcolor{black}{525.7} & \textcolor{black}{524.0} & \textcolor{black}{513.1} & \textcolor{black}{Ti-3V} & \textcolor{black}{439.6} & \textcolor{black}{433.5} & \textcolor{black}{657.0} & \textcolor{black}{679.1} & \textcolor{black}{617.8} & \textcolor{black}{Ti-1.5Mo} & \textcolor{black}{759.5} & \textcolor{black}{755.5} & \textcolor{black}{913.8} & \textcolor{black}{910.1} & \textcolor{black}{871.7} \\
\textcolor{black}{Ti-8Al} & \textcolor{black}{555.8} & \textcolor{black}{556.0} & \textcolor{black}{599.6} & \textcolor{black}{598.0} & \textcolor{black}{587.1} & \textcolor{black}{Ti-4V} & \textcolor{black}{499.4} & \textcolor{black}{493.4} & \textcolor{black}{716.9} & \textcolor{black}{739.0} & \textcolor{black}{677.7} & \textcolor{black}{Ti-2.0Mo} & \textcolor{black}{858.0} & \textcolor{black}{854.0} & \textcolor{black}{1012.3} & \textcolor{black}{1008.6} & \textcolor{black}{970.2} \\
\bottomrule
\end{tabular}%
}
\end{table}

Comparing our methodology with the existing one, the EET-based approach~\citep{huang2023strengthening} generally neglects SRO effects. As a result, EET-based methods incorrectly identify Al-Al bonds as the dominant strengthening interaction in Ti-Al systems. This misidentification arises from EET's inability to account for local coordination preferences. Our first-principles approach identifies the SRO (ruling out first near neighbor Al-Al bonds) and directly computes bond strengths from electronic structure, providing a more realistic representation of atomic-scale environments. This underscores the importance of DFT-derived electronic descriptors over the empirical parameters for predictive materials design.




\section{Conclusion}
This work has successfully developed a first-principles-based framework for predicting the tensile strength of $\alpha$-titanium alloys. Through a comprehensive investigation combining large-scale Density Functional Theory (DFT) calculations with advanced quantum-chemical bonding analysis, we have drawn several key conclusions:

(i) \textit{Short-Range Ordering is a Dominant Factor:} We have demonstrated that common substitutional solutes (Al, V, Mo) in $\alpha$-Ti exhibit a strong and consistent energetic preference for 2NN SRO. Interstitial solutes play a complex, chemistry-dependent role. For example, O prefers to stay at octahedral voids formed by pure-Ti, avoiding mixed Ti-Al octahedral voids. On the other hand, C prefers to stay at mixed Ti-V octahedral voids. A physically accurate description of these alloys must account for these non-random local atomic arrangements.

(ii) \textit{ICOHP is a Robust Descriptor for Bond Strength:} The Integrated Crystal Orbital Hamilton Population (ICOHP), derived from a projection of the DFT wavefunctions, serves as a direct, quantitative, and physically meaningful descriptor of covalent bond strength. Our analysis has established a clear hierarchy of bond strengths that mechanistically explains the relative strengthening of different alloying elements.

(iii) \textit{A \textcolor{black}{First-Principles-Informed Semi-Empirical} Tensile Strength Model:}
Building on the ICOHP descriptor, a tensile strength model is formulated that explicitly incorporates SRO effects and quantifies strengthening from both dislocation nucleation ($\gamma_{\text{ICOHP}}$) and glide resistance
($\Delta\eta_{\text{ICOHP}}$). \textcolor{black}{Unlike EET-based approaches, in which the bond-energy parameters themselves are empirically fitted, the bonding descriptors entering our model are obtained without adjustable parameters directly from the DFT electronic structure; the model framework of Eq.~\ref{eq:A1} is adopted from Ref.~\citep{huang2023strengthening}, and the concentration weights retain two empirically fitted coefficients. The model is therefore best described as a first-principles-informed semi-empirical model rather than a purely first-principles one. It predicts} tensile strengths of \textcolor{black}{$\alpha$-Ti alloys with mean deviations of 6.1\% (Ti-Al) and 2.7\% (Ti-Mo) from experimental data, using a single calibration constant per experimentally calibrated binary system; the interstitial contributions require no additional parameter.}

The primary novelty of this work lies in the successful paradigm shift from correlational, empirical parameters to quantum-chemical descriptors. By building a model upon the fundamental physics of chemical bonding and explicitly incorporating the effects of local atomic order, we have created an extensible platform for the computational design of materials. This framework represents a significant advancement in our ability to rationally engineer the next generation of high-performance titanium alloys.

\section{Acknowledgments} We acknowledge the National Super Computing Mission (NSM) for providing computing resources of ``PARAM Sanganak'' at IIT Kanpur, which is implemented by CDAC and supported by the Ministry of Electronics and Information Technology (MeitY) and the Department of Science and Technology (DST), Government of India. We also thank the ICME National Hub, IIT Kanpur, and CC, IIT Kanpur, for providing an HPC facility. Md. Faiz Akhtar acknowledges financial support from the Prime Minister's Research Fellows (PMRF) scheme.

\section{Data availability}
Data will be made available on request.

\appendix
\section{Computational Procedures for Strength Increments}
The tensile strength of an $\alpha$-type titanium alloy is calculated following the methodology of \citep{huang2023strengthening} as:
\begin{equation}
    \sigma_{\text{tensile}} = \sigma_0 + \sum \sigma_{\gamma} 
    + \sum \sigma_{\Delta\eta}
    \label{eq:A1}
\end{equation}
where $\sigma_0$ is the strength of pure $\alpha$-Ti, and $\sigma_{\gamma}$ and $\sigma_{\Delta\eta}$ are the strength increments arising from lattice resistance to dislocation nucleation and gliding, respectively. These are expressed as:
\begin{equation}
    \sigma_{\gamma} = \sigma_0 \cdot W_{\gamma} \cdot 
    \gamma_{\text{ICOHP}}, \qquad
    \sigma_{\Delta\eta} = \sigma_0 \cdot W_{\Delta\eta} \cdot 
    \Delta\eta_{\text{ICOHP}}
    \label{eq:A2}
\end{equation}
where $\gamma_{\text{ICOHP}}$ and $\Delta\eta_{\text{ICOHP}}$ are the ICOHP-derived bond-strength and interfacial 
coefficients defined in Section~3.4, and $W_{\gamma}$, $W_{\Delta\eta}$ are concentration-dependent weights given by:
\begin{equation}
    W_{\gamma} = \frac{C \times 100}{p}, \qquad
    W_{\Delta\eta} = \frac{C \times 100}{q}
    \label{eq:A3}
\end{equation}
where $C$ is the atomic concentration of the alloying element, and $p$ and $q$ are distribution and segregation coefficients, respectively, fitted to experimental tensile strength data. For ternary Ti-X-Y alloys, the total tensile strength is obtained by summing the binary Ti-X contribution with the additional increment introduced by the interstitial Y, using the thermodynamically preferred ICOHP values identified from the binding energy analysis (Table~\ref{tab:binding_energies_heatmap}).

\textcolor{black}{For clarity, we state explicitly which elements of this framework are inherited and which are introduced here. The overall additive structure of Eq.~\ref{eq:A1}, the decomposition of the strength increment into a nucleation term and a gliding term in Eq.~\ref{eq:A2}, and the functional form of the concentration weights in Eq.~\ref{eq:A3} are adopted unchanged from Huang \textit{et al.}~\citep{huang2023strengthening}. What is new in the present work is the content of the two coefficients: in place of the empirically parameterized valence-electron descriptors of the Empirical Electron Theory, we substitute $\gamma_{\text{ICOHP}}$ and $\Delta\eta_{\text{ICOHP}}$, which are computed without adjustable parameters directly from the DFT electronic structure (Eqs.~\ref{eq:gamma_icohp}-\ref{eq:delta_eta_icohp}), and which are evaluated at the thermodynamically preferred atomic configuration identified by the binding-energy analysis rather than at an assumed random solid solution. The coefficients $p$ and $q$ remain empirical: they are not computed from first principles but are obtained by least-squares fitting of Eq.~\ref{eq:A1} to experimental tensile-strength data for the corresponding binary alloy series. Accordingly, the model is a first-principles-informed semi-empirical model, in which the bonding physics is parameter-free but the concentration scaling is calibrated against experiment.}

\textcolor{black}{Substituting Eqs.~\ref{eq:A2} and~\ref{eq:A3} into Eq.~\ref{eq:A1}, $p$ and $q$ enter only through the combination $\gamma_{\text{ICOHP}}/p + \Delta\eta_{\text{ICOHP}}/q$ and are not separately identifiable; we therefore set $p = q$ and report one calibration constant per binary system. $C$ is an atomic percentage, so compositions reported in weight percent are converted first: the 2.02, 3.90, 5.80 and 7.80~wt.\% Al alloys of Ref.~\citep{youssef2021effects} correspond to 3.53, 6.72, 9.85 and 13.06~at.\%, while Ref.~\citep{kobayashi2024effects} is already in at.\%. Using $|\text{ICOHP}|_{Ti} = 0.76$~eV with the values of Table~\ref{tab:icohp_ternary_manual}, $\gamma$ and $\Delta\eta$ are 0.449 and 0.366 for Ti-Al, 0.118 and 0.112 for Ti-V, and 0.695 and 0.516 for Ti-Mo. Least-squares fitting of Eq.~\ref{eq:A1} with the intercept held at $\sigma_0$ then gives $p = q = 5.73$ for Ti-Al ($\sigma_0 = 260$~MPa~\citep{huang2023strengthening}) and $2.85$ for Ti-Mo ($\sigma_0 = 464$~MPa, measured in the same study~\citep{kobayashi2024effects}), with mean deviations of 6.1\% and 2.7\% relative to experiment.}



\bibliographystyle{elsarticle-num}
\bibliography{reference.bib}

@article{williams2020opportunities,
  title={Opportunities and issues in the application of titanium alloys for aerospace components},
  author={Williams, James C and Boyer, Rodney R},
  journal={Metals},
  volume={10},
  number={6},
  pages={705},
  year={2020},
  publisher={MDPI}, 
  doi = {https://www.mdpi.com/2075-4701/10/6/705}
}

@article{boyer1996overview,
  title={An overview on the use of titanium in the aerospace industry},
  author={Boyer, Renee R},
  journal={Materials Science and Engineering: A},
  volume={213},
  number={1-2},
  pages={103--114},
  year={1996},
  publisher={Elsevier}, 
  doi = {https://www.sciencedirect.com/science/article/pii/0921509396102331}
}

@article{kolli2018review,
  title={A review of metastable beta titanium alloys},
  author={Kolli, R Prakash and Devaraj, Arun},
  journal={Metals},
  volume={8},
  number={7},
  pages={506},
  year={2018},
  publisher={MDPI}, 
  doi = {https://www.mdpi.com/2075-4701/8/7/506}
}

@article{yu2015origin,
  title={Origin of dramatic oxygen solute strengthening effect in titanium},
  author={Yu, Qian and Qi, Liang and Tsuru, Tomohito and Traylor, Rachel and Rugg, David and Morris Jr, JW and Asta, Mark and Chrzan, DC and Minor, Andrew M},
  journal={Science},
  volume={347},
  number={6222},
  pages={635--639},
  year={2015},
  publisher={American Association for the Advancement of Science}, 
  doi =  {https://www.science.org/doi/10.1126/science.1260485}
}

@article{zhang2019direct,
  title={Direct imaging of short-range order and its impact on deformation in Ti-6Al},
  author={Zhang, Ruopeng and Zhao, Shiteng and Ophus, Colin and Deng, Yu and Vachhani, Shraddha J and Ozdol, Burak and Traylor, Rachel and Bustillo, Karen C and Morris Jr, JW and Chrzan, Daryl C and others},
  journal={Science advances},
  volume={5},
  number={12},
  pages={eaax2799},
  year={2019},
  publisher={American Association for the Advancement of Science}, 
  doi = {https://www.science.org/doi/full/10.1126/sciadv.aax2799}
}

@article{li2023quantitative,
  title={Quantitative three-dimensional imaging of chemical short-range order via machine learning enhanced atom probe tomography},
  author={Li, Yue and Wei, Ye and Wang, Zhangwei and Liu, Xiaochun and Colnaghi, Timoteo and Han, Liuliu and Rao, Ziyuan and Zhou, Xuyang and Huber, Liam and Dsouza, Raynol and others},
  journal={Nature Communications},
  volume={14},
  number={1},
  pages={7410},
  year={2023},
  publisher={Nature Publishing Group UK London}, 
  doi = {https://www.nature.com/articles/s41467-023-43314-y}
}

@article{he2024quantifying,
  title={Quantifying short-range order using atom probe tomography},
  author={He, Mengwei and Davids, William J and Breen, Andrew J and Ringer, Simon P},
  journal={Nature Materials},
  volume={23},
  number={9},
  pages={1200--1207},
  year={2024},
  publisher={Nature Publishing Group UK London}, 
  doi = {https://www.nature.com/articles/s41563-024-01912-1}
}

@article{kwasniak2023competition,
  title={Competition between prismatic and basal slip in hexagonal titanium--aluminum alloys with short-range order},
  author={Kwasniak, Piotr and Delannoy, St{\'e}phanie and Prima, Fr{\'e}d{\'e}ric and Clouet, Emmanuel},
  journal={Materials Research Letters},
  volume={11},
  number={6},
  pages={407--413},
  year={2023},
  publisher={Taylor \& Francis}, 
  doi = {https://www.tandfonline.com/doi/full/10.1080/21663831.2023.2169082}
}

@article{yin2017comprehensive,
  title={Comprehensive first-principles study of stable stacking faults in hcp metals},
  author={Yin, Binglun and Wu, Zhaoxuan and Curtin, WA},
  journal={Acta Materialia},
  volume={123},
  pages={223--234},
  year={2017},
  publisher={Elsevier}, 
  doi = {https://www.sciencedirect.com/science/article/pii/S1359645416308035}
}

@article{calazans2024recent,
  title={Recent advances and prospects in $\beta$-type titanium alloys for dental implants applications},
  author={Calazans Neto, João V and Celles, C{\'\i}cero AS and de Andrade, Catia SAF and Afonso, Conrado RM and Nagay, Bruna E and Bar{\~a}o, Valentim AR},
  journal={ACS Biomaterials Science \& Engineering},
  volume={10},
  number={10},
  pages={6029--6060},
  year={2024},
  publisher={ACS Publications}, 
  doi = {https://pubs.acs.org/doi/full/10.1021/acsbiomaterials.4c00963}
}

@article{ren2025effect,
  title={Effect of Interstitial Oxygen on the Microstructure and Mechanical Properties of Titanium Alloys: A Review},
  author={Ren, Yaojia and Xu, Jiajun and Wei, Yingkang and Liu, Yingying and Zhu, Jilei and Liu, Shifeng},
  journal={Crystals},
  volume={15},
  number={7},
  pages={618},
  year={2025},
  publisher={MDPI}, 
  doi = {https://www.mdpi.com/2073-4352/15/7/618}
}

@article{ghazisaeidi2014interaction,
  title={Interaction of oxygen interstitials with lattice faults in Ti},
  author={Ghazisaeidi, M and Trinkle, DR},
  journal={Acta Materialia},
  volume={76},
  pages={82--86},
  year={2014},
  publisher={Elsevier}, 
  doi = {https://doi.org/10.1016/j.actamat.2014.05.025}
}

@article{bakulin2022impurity,
  title={Impurity combination effect on oxygen absorption in $\alpha$2-Ti3Al},
  author={Bakulin, Alexander V and Chumakova, Lora S and Kasparyan, Sergey O and Kulkova, Svetlana E},
  journal={Metals},
  volume={12},
  number={4},
  pages={650},
  year={2022},
  publisher={MDPI}, 
  doi = {https://doi.org/10.3390/met12040650}
}

@article{bakulin2020diffusion,
  title={Diffusion properties of oxygen in the $\gamma$-TiAl alloy},
  author={Bakulin, AV and Kulkov, SS and Kulkova, SE},
  journal={Journal of Experimental and Theoretical Physics},
  volume={130},
  number={4},
  pages={579--590},
  year={2020},
  publisher={Springer}, 
  doi = {https://link.springer.com/article/10.1134/S1063776120030115}
}

@article{ghosh2021effect,
  title={Effect of oxygen interstitials on structural stability in refractory metals (V, Mo, W) from DFT calculations},
  author={Ghosh, Sutapa and Ghosh, Chanchal},
  journal={The European Physical Journal B},
  volume={94},
  number={5},
  pages={114},
  year={2021},
  publisher={Springer}, 
doi = {https://link.springer.com/article/10.1140/epjb/s10051-021-00110-1}
}

@article{huang2023strengthening,
  title={Strengthening effects of Al element on strength and impact toughness in titanium alloy},
  author={Huang, Shixing and Zhao, Qinyang and Yang, Zhong and Lin, Cheng and Zhao, Yongqing and Yu, Jiashi},
  journal={Journal of Materials Research and Technology},
  volume={26},
  pages={504--516},
  year={2023},
  publisher={Elsevier}, 
  doi = {https://doi.org/10.1016/j.jmrt.2023.07.206}
}

@article{qu2011theoretical,
  title={Theoretical Calculation of $\beta$ Transition Temperature of Ti-6Al-4V from Valence Electron Level},
  author={Qu, Hua and Liu, Wei Dong},
  journal={Advanced Materials Research},
  volume={299},
  pages={592--595},
  year={2011},
  publisher={Trans Tech Publ}, 
  doi = {https://www.scientific.net/AMR.299-300.592}
}

@article{lin2011analysis,
  title={Analysis of the effect of alloy elements on martensitic transformation in titanium alloy with the use of valence electron structure parameters},
  author={Lin, Cheng and Yin, Guili and Zhao, Yongqing and Ge, Peng and Liu, Zhilin},
  journal={Materials Chemistry and Physics},
  volume={125},
  number={3},
  pages={411--417},
  year={2011},
  publisher={Elsevier}, 
  doi = {https://www.sciencedirect.com/science/article/pii/S0254058410008709}
}

@misc{maintz2016lobster,
  title={LOBSTER: a tool to extract chemical bonding from plane-wave based DFT},
  author={Maintz, Stefan and Deringer, Volker L and Tchougr{\'e}eff, Andrei L and Dronskowski, Richard},
  year={2016},
  publisher={Wiley Online Library}, 
  doi = {https://onlinelibrary.wiley.com/doi/full/10.1002/jcc.24300}
}

@article{lin2016simple,
  title={Simple models to account for the formation and decomposition of athermal $\omega$ phase in titanium alloys},
  author={Lin, Cheng and Yin, Guili and Zhang, Aimin and Zhao, Yongqing and Li, Qingchun},
  journal={Scripta Materialia},
  volume={117},
  pages={28--31},
  year={2016},
  publisher={Elsevier}, 
  doi = {https://doi.org/10.1016/j.scriptamat.2016.01.042}
}

@article{dronskowski1993crystal,
  title={Crystal orbital Hamilton populations (COHP): energy-resolved visualization of chemical bonding in solids based on density-functional calculations},
  author={Dronskowski, Richard and Bloechl, Peter E},
  journal={The Journal of Physical Chemistry},
  volume={97},
  number={33},
  pages={8617--8624},
  year={1993},
  publisher={ACS Publications}, 
  doi = {https://doi.org/10.1021/j100135a014}
}

@article{deringer2011crystal,
  title={Crystal orbital Hamilton population (COHP) analysis as projected from plane-wave basis sets},
  author={Deringer, Volker L and Tchougr{\'e}eff, Andrei L and Dronskowski, Richard},
  journal={The journal of physical chemistry A},
  volume={115},
  number={21},
  pages={5461--5466},
  year={2011},
  publisher={ACS Publications}, 
  doi = {https://pubs.acs.org/doi/10.1021/jp202489s}
}

@article{steinberg2018crystal,
  title={The crystal orbital Hamilton population (COHP) method as a tool to visualize and analyze chemical bonding in intermetallic compounds},
  author={Steinberg, Simon and Dronskowski, Richard},
  journal={Crystals},
  volume={8},
  number={5},
  pages={225},
  year={2018},
  publisher={MDPI}, 
  doi = {https://doi.org/10.3390/cryst8050225}
}

@article{muller2021crystal,
  title={Crystal orbital bond index: covalent bond orders in solids},
  author={M\"uller, Peter C and Ertural, Christina and Hempelmann, Jan and Dronskowski, Richard},
  journal={The Journal of Physical Chemistry C},
  volume={125},
  number={14},
  pages={7959--7970},
  year={2021},
  publisher={ACS Publications}, 
  doi = {https://pubs.acs.org/doi/full/10.1021/acs.jpcc.1c00718}
}

@article{das2024interplay,
  title={The interplay of chemical bonding and thermoelectric properties in doped cubic GeTe},
  author={Das, Sree Sourav and Sadeghi, Safoura Nayeb and Esfarjani, Keivan and Zebarjadi, Mona},
  journal={Journal of Materials Chemistry A},
  volume={12},
  number={23},
  pages={14072--14086},
  year={2024},
  publisher={Royal Society of Chemistry}, 
  doi = {https://pubs.rsc.org/en/content/articlehtml/2024/ta/d4ta01088d}
}

@article{al2024accelerating,
  title={Accelerating Discovery of Extreme Lattice Thermal Conductivity by Crystal Attention Graph Neural Network (CATGNN) Using Chemical Bonding Intuitive Descriptors},
  author={Al-Fahdi, Mohammed and Rurali, Riccardo and Hu, Jianjun and Wolverton, Christopher and Hu, Ming},
  journal={arXiv preprint arXiv:2410.16066},
  year={2024}, 
  doi = {https://arxiv.org/abs/2410.16066}
}

@article{youssef2021effects,
  title={Effects of Al content and $\alpha$2 precipitation on the fatigue crack growth behaviors of binary Ti--Al alloys},
  author={Youssef, Sabry S and Zheng, Xiaodong and Qi, Min and Ma, Yingjie and Huang, Sensen and Qiu, Jianke and Zheng, Shijian and Lei, Jiafeng and Yang, Rui},
  journal={Materials Science and Engineering: A},
  volume={819},
  pages={141513},
  year={2021},
  publisher={Elsevier}, 
  doi = {https://www.sciencedirect.com/science/article/pii/S0921509321007826}
}

@article{kobayashi2024effects,
  title={The effects of oxygen addition on microstructure and mechanical properties of Ti-Mo alloys for biomedical application},
  author={Kobayashi, Sengo and Okano, Satoshi},
  journal={Frontiers in Bioengineering and Biotechnology},
  volume={12},
  pages={1380503},
  year={2024},
  publisher={Frontiers Media SA}, 
  doi = {https://doi.org/10.3389/fbioe.2024.1380503}
}

@article{faiz2025novel,
  title={A novel electronic structure-based prediction of interstitial strengthening in $\alpha$-titanium},
  author={Faiz Akhtar, Md and Gurao, Nilesh P and Bhowmick, Somnath},
  journal={Journal of Applied Physics},
  volume={138},
  number={17},
  year={2025},
  publisher={AIP Publishing}, 
  doi = {https://pubs.aip.org/aip/jap/article/138/17/175111/3371105}
}

@article{chong2020mechanistic,
  title={Mechanistic basis of oxygen sensitivity in titanium},
  author={Chong, Yan and Poschmann, Max and Zhang, Ruopeng and Zhao, Shiteng and Hooshmand, Mohammad S and Rothchild, Eric and Olmsted, David L and Morris Jr, JW and Chrzan, Daryl C and Asta, Mark and others},
  journal={Science advances},
  volume={6},
  number={43},
  pages={eabc4060},
  year={2020},
  publisher={American Association for the Advancement of Science}, 
  url = {https://www.science.org/doi/10.1126/sciadv.abc4060}
}

@article{lindwall2018diffusion,
  title={Diffusion in the Ti-Al-V system},
  author={Lindwall, Greta and Moon, Kil-Won and Chen, Zhangqi and Mengason, Michael and Williams, Maureen E and Gorham, Justin M and Zhao, Ji-Cheng and Campbell, Carelyn E},
  journal={Journal of Phase Equilibria and Diffusion},
  volume={39},
  number={5},
  pages={731--746},
  year={2018},
  publisher={Springer}, 
  doi = {https://link.springer.com/article/10.1007/s11669-018-0673-9}
}

@article{wu2025observation,
  title={Observation of short-range order in refractory high-entropy alloys from atomic-positions deviation using STEM and atomistic simulations},
  author={Wu, Chia-Yi and Kim, George and Chang, Yuan-Wei and Li, Chenyang and Li, Juntan and Xu, Haixuan and Lee, Chanho and Liaw, Peter K and Chen, Wei and Chou, Yi-Chia},
  journal={Materials Today Physics},
  pages={101796},
  year={2025},
  publisher={Elsevier}, 
  doi = {https://www.sciencedirect.com/science/article/pii/S254252932500152X}
}

@article{hooshmand2018first,
  title={First-principles prediction of oxygen diffusivity near the (101{\={}} 2) twin boundary in titanium},
  author={Hooshmand, MS and Niu, C and Trinkle, DR and Ghazisaeidi, M},
  journal={Acta Materialia},
  volume={156},
  pages={11--19},
  year={2018},
  publisher={Elsevier}, 
  doi = {https://www.sciencedirect.com/science/article/pii/S1359645418304762}
}

@article{gunda2020understanding,
  title={Understanding the interactions between interstitial and substitutional solutes in refractory alloys: The case of Ti-Al-O},
  author={Gunda, NS Harsha and Van der Ven, Anton},
  journal={Acta Materialia},
  volume={191},
  pages={149--157},
  year={2020},
  publisher={Elsevier},
  doi = {https://www.sciencedirect.com/science/article/pii/S1359645420302743}
}

@article{rohling2019correlations,
  title={Correlations between density-based bond orders and orbital-based bond energies for chemical bonding analysis},
  author={Rohling, Roderigh Y and Tranca, Ionut C and Hensen, Emiel JM and Pidko, Evgeny A},
  journal={The Journal of Physical Chemistry C},
  volume={123},
  number={5},
  pages={2843--2854},
  year={2019},
  publisher={ACS Publications}, 
  doi = {https://pubs.acs.org/doi/full/10.1021/acs.jpcc.8b08934}
}

@article{PhysRevB.59.1758,
  author = {Kresse, G. and Joubert, D.},
  title = {From ultrasoft pseudopotentials to the projector augmented-wave method},
  journal = {Phys. Rev. B},
  volume = {59},
  issue = {3},
  pages = {1758--1775},
  numpages = {0},
  year = {1999},
  month = {Jan},
  publisher = {American Physical Society},
  doi = {10.1103/PhysRevB.59.1758},
  url = {https://link.aps.org/doi/10.1103/PhysRevB.59.1758}
}

@article{blochl1994projector,
  title={Projector augmented-wave method},
  author={Bl{\"o}chl, Peter E},
  journal={Physical review B},
  volume={50},
  number={24},
  pages={17953},
  year={1994},
  publisher={APS},
  url = {https://journals.aps.org/prb/abstract/10.1103/PhysRevB.50.17953}
}

@article{PhysRevB.54.11169,
  author = {Kresse, G. and Furthm\"uller, J.},
  title = {Efficient iterative schemes for ab initio total-energy calculations using a plane-wave basis set},
  journal = {Phys. Rev. B},
  volume = {54},
  issue = {16},
  pages = {11169--11186},
  numpages = {0},
  year = {1996},
  month = {Oct},
  publisher = {American Physical Society},
  doi = {10.1103/PhysRevB.54.11169},
  url = {https://link.aps.org/doi/10.1103/PhysRevB.54.11169}
}

@article{PhysRevLett.77.3865,
  author = {Perdew, John P. and others},
  title= {Generalized Gradient Approximation Made Simple},
  journal = {Phys. Rev. Lett.},
  volume = {77},
  issue = {18},
  pages = {3865--3868},
  numpages = {0},
  year = {1996},
  month = {Oct},
  publisher = {American Physical Society},
  doi = {10.1103/PhysRevLett.77.3865},
  url = {https://link.aps.org/doi/10.1103/PhysRevLett.77.3865}
}

@article{monkhorst1976special,
  author={Monkhorst, Hendrik J and Pack, James D},
  title={Special points for Brillouin-zone integrations},
  journal={Phy. rev. B},
  volume={13},
  number={12},
  pages={5188},
  year={1976},
  publisher={APS}, 
url = {https://journals.aps.org/prb/abstract/10.1103/PhysRevB.13.5188}
}

@article{blochl1994improved,
  title={Improved tetrahedron method for Brillouin-zone integrations},
  author={Bl{\"o}chl, Peter E and Jepsen, Ove and Andersen, Ole Krogh},
  journal={Physical Review B},
  volume={49},
  number={23},
  pages={16223},
  year={1994},
  publisher={APS}, 
  doi = {https://journals.aps.org/prb/abstract/10.1103/PhysRevB.49.16223}
}

@article{feng2023microstructure,
  title={Microstructure and tensile properties of a multi-alloyed $\alpha$+ $\beta$ titanium alloy Ti4. 5Al10. 5V3Fe},
  author={Feng, Qisheng and Duan, Baohua and Jiao, Lina and Chen, Guangyao and Zou, Xingli and Lu, Xionggang and Li, Chonghe},
  journal={Materials Chemistry and Physics},
  volume={295},
  pages={127110},
  year={2023},
  publisher={Elsevier}, 
  doi = {https://www.sciencedirect.com/science/article/pii/S025405842201416X}
}

@article{burvsik1999ordering,
  title={Ordering of Substoichiometric $\delta$-TiCx Phase in Ti--V--C Alloys},
  author={Bur{\v{s}}{\'\i}k, J and Weatherly, GC},
  journal={physica status solidi (a)},
  volume={174},
  number={2},
  pages={327--335},
  year={1999},
  publisher={Wiley Online Library}, 
}

@article{bandyopadhyay2000ti,
  title={The Ti-VC system (titanium-vanadium-carbon)},
  journal={Journal of Phase Equilibria },
  author={Bandyopadhyay, Debashis},
  year={2000},
  publisher={Springer}, 
  doi = {https://link.springer.com/article/10.1361/105497100770340282}
}

@article{gusev1989short,
  title={Short-Range Order in Nonstoichiometrie Transition Metal Carbides, Nitrides, and Oxides},
  author={Gusev, AI},
  journal={physica status solidi (b)},
  volume={156},
  number={1},
  pages={11--40},
  year={1989},
  publisher={Wiley Online Library}, 
  doi = { https://doi.org/10.1002/pssb.2221560102}
}

@article{prysyazhnyuk2025first,
  title={First-Principles Study of the Mechanical Properties of (Ti, V) C Solid Solutions},
  author={Prysyazhnyuk, PM and Yaremiy, IP and Kharlov, AH and Makohin, MP and Umantsiv, IM and Savchyn, VV and Misiuk, OI},
  journal={Physics and Chemistry of Solid State},
  volume={26},
  number={2},
  pages={377--385},
  year={2025}, 
  doi = {https://journals.pnu.edu.ua/index.php/pcss/article/view/9092/9265}
}

@article{wang2022elastic,
  title={Elastic properties of solid-solution refractory metal carbides with vacancy from virtual crystal approximation and supercell method},
  author={Wang, Qingchun and Li, Qingrui and Ding, Hongsheng and Tian, Fuyang},
  journal={Computational Condensed Matter},
  volume={32},
  pages={e00721},
  year={2022},
  publisher={Elsevier},
  doi = {https://doi.org/10.1016/j.cocom.2022.e00721}
}

@article{vitek1968intrinsic,
  author={Vitek, Vaclav},
  title={Intrinsic stacking faults in body-centred cubic crystals},
  journal={Philosophical Magazine},
  volume={18},
  number={154},
  pages={773--786},
  year={1968},
  publisher={Taylor \& Francis},
  url = {https://www.tandfonline.com/doi/abs/10.1080/14786436808227500}
}

@article{lindaprb,
  title = {Accelerating the prediction of stacking fault energy by combining ab initio calculations and machine learning},
  author = {Linda, Albert and Akhtar, Md. Faiz and Pathak, Shaswat and Bhowmick, Somnath},
  journal = {Phys. Rev. B},
  volume = {109},
  issue = {21},
  pages = {214102},
  numpages = {11},
  year = {2024},
  month = {Jun},
  publisher = {American Physical Society},
  doi = {10.1103/PhysRevB.109.214102},
  url = {https://link.aps.org/doi/10.1103/PhysRevB.109.214102}
}

@article{muller2024fragment,
  title={Fragment Orbitals Extracted from First-Principles Plane-Wave Calculations},
  author={M\"uller, Peter C and Schmit, Nathalie and Sann, Leander and Steinberg, Simon and Dronskowski, Richard},
  journal={Inorganic Chemistry},
  volume={63},
  number={43},
  pages={20161--20172},
  year={2024},
  publisher={ACS Publications},
  doi = {https://doi.org/10.1021/acs.inorgchem.4c01024}
}

@article{reitz2024bonding,
  title={Bonding Analyses in the Broad Realm of Intermetallics: Understanding the Role of Chemical Bonding in the Design of Novel Materials},
  author={Reitz, Linda S and Hempelmann, Jan and M\"uller, Peter C and Dronskowski, Richard and Steinberg, Simon},
  journal={Chemistry of Materials},
  volume={36},
  number={14},
  pages={6791--6804},
  year={2024},
  publisher={ACS Publications},
  doi = {https://doi.org/10.1021/acs.chemmater.4c00425}
}

@article{momma2011vesta,
  title={{VESTA 3} for three-dimensional visualization of crystal, volumetric and morphology data},
  author={Momma, Koichi and Izumi, Fujio},
  journal={Journal of Applied Crystallography},
  volume={44},
  number={6},
  pages={1272--1276},
  year={2011},
  publisher={International Union of Crystallography},
  doi = {https://doi.org/10.1107/S0021889811038970}
}
\end{document}